\documentclass{article}

\usepackage[pdftex,breaklinks,colorlinks,
citecolor=blue,
urlcolor=blue]{hyperref}
\usepackage{float}
\usepackage{multicol}
\usepackage{subfigure}
\usepackage{tabularx}
\usepackage{booktabs}
\usepackage{graphicx}
\usepackage{amsmath}
\usepackage{amssymb}
\usepackage{latexsym}
\usepackage{mathrsfs}
\usepackage{geometry}
\usepackage{chngcntr}
\counterwithin{figure}{section}
\counterwithin{table}{section}
\numberwithin{equation}{section}
\DeclareMathOperator{\sech}{sech}
\usepackage{authblk}
\title{A second look at transition amplitudes in $(2+1)$-dimensional causal dynamical triangulations}
\author[1]{Joshua H. Cooperman}
\author[2]{Kyle Lee}
\author[3,4]{Jonah M. Miller}
\affil[1]{Physics Program, Bard College, Annandale-on-Hudson, New York, United States}
\affil[2]{C. N. Yang Institute for Theoretical Physics, State University of New York, Stony Brook, New York, United States}
\affil[3]{Perimeter Institute for Theoretical Physics, Waterloo, Ontario, Canada}
\affil[4]{Department of Physics, University of Guelph, Guelph, Ontario, Canada}
\begin{document}

\maketitle

\begin{abstract}
Studying transition amplitudes in $(2+1)$-dimensional causal dynamical triangulations, Cooperman and Miller discovered speculative evidence for Lorentzian quantum geometries emerging from its Euclidean path integral \cite{JHC&JMM}. On the basis of this evidence, Cooperman and Miller conjectured that Lorentzian de Sitter spacetime, not Euclidean de Sitter space, dominates the ground state of the quantum geometry of causal dynamical triangulations on large scales, a scenario akin to that of the Hartle-Hawking no-boundary proposal in which Lorentzian spacetimes dominate a Euclidean path integral \cite{JBH&SWH}. We argue against this conjecture: we propose a more straightforward explanation of their findings, and we proffer evidence for the Euclidean nature of these seemingly Lorentzian quantum geometries. This explanation reveals another manner in which 
the Euclidean path integral of causal dynamical triangulations behaves correctly in its semiclassical limit---the implementation and interaction of multiple constraints. 
\end{abstract}

\section{Euclidean from Lorentzian}\label{introduction}


One often studies a Poincar\'e-invariant quantum field theory defined on Minkowski spacetime \emph{via} Wick rotation to a Euclidean-invariant statistical field theory defined on Euclidean space. Within the path integral formulation, the Wick rotation transforms a Lorentzian path integral, which involves complex probability amplitudes for each Lorentzian field configuration, into a statistical partition function, which involves real probabilities for each Euclidean field configuration. Absent the complications of complex probability amplitudes for Lorentzian field configurations, calculations typically prove considerably more tractable. 
Provided that the statistical field theory satisfies the Osterwalder-Schrader axioms, one can recover the Lorentzian theory from the Euclidean theory through the Osterwalder-Schrader reconstruction theorem \cite{KO&RS1,KO&RS2}. One thus defines the Lorentzian theory in terms of the Euclidean theory. 

The tempting prospect that a quantum theory of gravity could be similarly defined led to the development of various approaches taking as their starting point the partition function
\begin{equation}\label{formalEpathintegral}
\mathscr{Z}[\mathbf{\gamma}]=\int_{\mathbf{g}|_{\partial\mathcal{M}}=\mathbf{\gamma}}\mathrm{d}\mu(\mathbf{g})\,e^{-S_{\mathrm{cl}}^{(\mathrm{E})}[\mathbf{g}]/\hbar}
\end{equation}
over Euclidean geometries specified by a metric tensor $\mathbf{g}$. 
One should, however, be skeptical of these approaches' applicability to gravity: a typical spacetime, even satisfying the Einstein equations, does not permit a global Wick rotation from Lorentzian to Euclidean signature. 
Nevertheless, such approaches---collectively called Euclidean quantum gravity---work not only sensibly, but even successfully in sundry circumstances \cite{GWG&SWH2}. We briefly mention two notable examples. First, one can derive the thermodynamic behavior of black holes from the partition function \eqref{formalEpathintegral}. Gibbons and Hawking computed the black hole entropy \cite{GWG&SWH}, and Hartle and Hawking computed the black hole radiance \cite{JBH&SWH1}. Second, Hartle and Hawking developed a quantum theory of gravity in the minisuperspace truncation from the partition function \eqref{formalEpathintegral}, their so-called no-boundary proposal \cite{JBH&SWH}. These authors defined a wave function for the universe having a remarkable property: 
Lorentzian geometries dominate the partition function \eqref{formalEpathintegral} owing to 
the necessity of deforming an integration contour into the complex plane. 
Consequently, there is no need for an Osterwalder-Schrader reconstruction: the partition function \eqref{formalEpathintegral} directly defines a Lorentzian quantum theory of gravity. Initial attempts to construct a complete nonperturbative quantum theory of gravity on the basis of the partition function \eqref{formalEpathintegral} did not fare so well \cite{RL}. Two approaches, quantum Regge calculus and Euclidean dynamical triangulations, both grounded upon lattice regularization of the partition function \eqref{formalEpathintegral}, were extensively studied \cite{RL}. Neither of the quantum theories of gravity so defined exhibited a sufficiently rich phase structure to support a continuum limit.\footnote{See \cite{HWH}, however.} More recently, an approach based on exact renormalization group analysis of the partition function \eqref{formalEpathintegral} has shown promise \cite{MR&FS}.

Causal dynamical triangulations emerged from the failures of quantum Regge calculus and Euclidean dynamical triangulations \cite{JA&RL}. This newer approach takes as its starting point the Lorentzian path integral
\begin{equation}
\mathscr{A}[\gamma]=\int_{\mathbf{g}|_{\partial\mathcal{M}}=\gamma}\mathrm{d}\mu(\mathbf{g})\,e^{iS_{\mathrm{cl}}[\mathbf{g}]/\hbar}
\end{equation}
over Lorentzian geometries specified by a metric tensor $\mathbf{g}$. One chooses to restrict the path integration to appropriately causal Lorentzian geometries, namely, those admitting a global foliation by spacelike hypersurfaces all of fixed topology. 
One then introduces a lattice regularization---causal triangulations---of these causal Lorentzian geometries. 
As Ambj\o rn, Jurkiewicz, and Loll demonstrated, this restriction allows for a well-defined Wick rotation of any Lorentzian causal triangulation to a corresponding Euclidean causal triangulation \cite{JA&JJ&RL1,JA&JJ&RL2}. This Wick rotation enables the use of Monte Carlo methods to study the resulting partition function. 

Having implemented this Wick rotation, one could have wondered if the resulting partition function behaves conventionally, such as that of a 
field theory satisfying the Osterwalder-Schrader axioms, or unconventionally, such as that of the Hartle-Hawking no-boundary proposal. On the basis of Monte Carlo simulations of certain transition amplitudes within the causal dynamical triangulations of $(2+1)$-dimensional Einstein gravity, Cooperman and Miller conjectured that its partition function 
behaves unconventionally \cite{JHC&JMM}. Specifically, these authors suggested that geometries resembling Lorentzian de Sitter spacetime---not, as previously thought, Euclidean de Sitter space---on sufficiently large scales dominate this partition function. Independently, Ambj\o rn \emph{et al} argued for a signature change transition within the causal dynamical triangulations of $(3+1)$-dimensional Einstein gravity \cite{JA&DNC&JGS&JJ}. We now argue, contrary to the conjecture of Cooperman and Miller, that the partition function of causal dynamical triangulations behaves conventionally. Specifically, by reinterpreting these Monte Carlo simulations, we maintain that geometries resembling Euclidean de Sitter space on sufficiently large scales indeed dominate this partition function. In the process of making this argument, we provide further evidence that the partition function of causal dynamical triangulations behaves correctly in its semiclassical limit. 

We introduce the formalism of causal dynamical triangulations, specializing to the case of $2+1$ dimensions for spherical topology with initial and final spacelike boundaries, in section \ref{CDT}. After recalling the relevant results from \cite{JHC&JMM} and presenting new related results, we restate the conjecture of Cooperman and Miller in section \ref{evidenceconjecture}. We present a first analysis of all of these results in section \ref{analysissupport}, which offers evidence in support of their conjecture. We present a more careful analysis in section \ref{argumentrefutation}, which leads to our argument refuting the conjecture of Cooperman and Miller. We conclude in section \ref{conclusion} by echoing Cooperman's call for the proof of an Osterwalder-Schrader-type theorem for causal dynamical triangulations \cite{JHC2}. Four appendices supplement aspects of sections \ref{CDT}, \ref{analysissupport}, and \ref{argumentrefutation}.

\section{Causal dynamical triangulations}\label{CDT}

Within a path integral quantization of a classical metric theory of gravity, one formally defines a transition amplitude as 
\begin{equation}\label{gravitypathintegral}
\mathscr{A}[\gamma]=\int_{\mathbf{g}|_{\partial\mathcal{M}}=\gamma}\mathrm{d}\mu(\mathbf{g})\,e^{iS_{\mathrm{cl}}[\mathbf{g}]/\hbar}.
\end{equation}
The right hand side of equation \eqref{gravitypathintegral} encodes the following instructions for computing the transition amplitude $\mathscr{A}[\gamma]$: integrate over all spacetime metric tensors $\mathbf{g}$ that induce the metric tensor $\gamma$ on the boundary $\partial\mathcal{M}$ of the spacetime manifold $\mathcal{M}$, 
weighting each metric tensor $\mathbf{g}$ by the product of the measure $\mathrm{d}\mu(\mathbf{g})$ and the exponential $e^{iS_{\mathrm{cl}}[\mathbf{g}]/\hbar}$. $S_{\mathrm{cl}}[\mathbf{g}]$ is the action specifying the classical metric theory of gravity, including boundary terms enforcing the condition $\mathbf{g}|_{\partial\mathcal{M}}=\gamma$. 

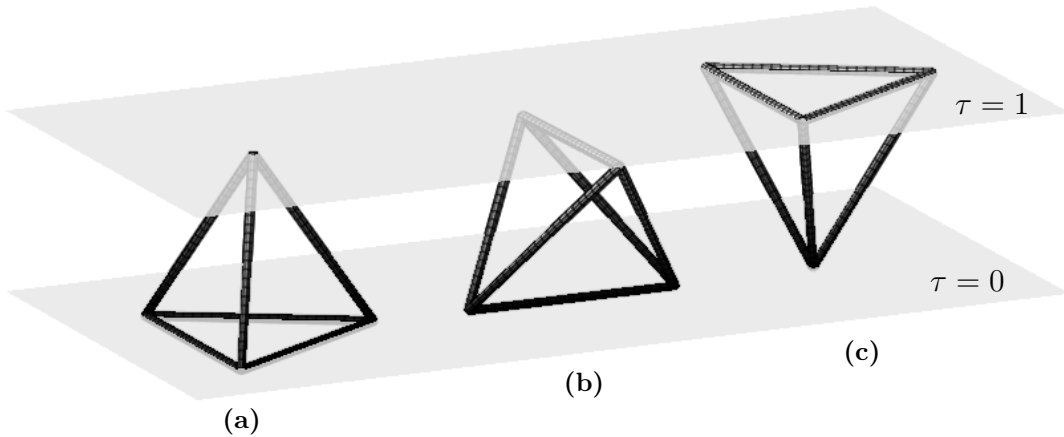
\begin{figure}
\centering
\setlength{\unitlength}{\textwidth}
\begin{picture}(1,0.25)
\put(0,0){\includegraphics[clip,scale=0.5,trim={0cm 1.7cm 0cm 0cm}]{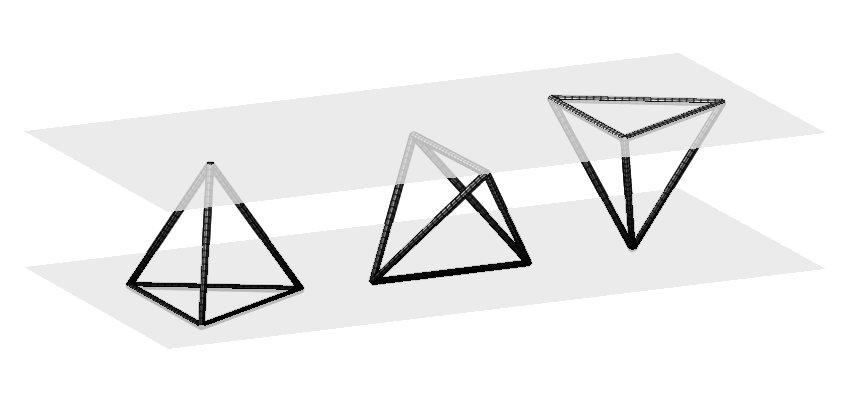}}
\put(0.2,-0.02){\bf (a)}
\put(0.475,0.01){\bf (b)}
\put(0.7,0.035){\bf (c)}
\put(0.77,0.09){\large $\tau = 0$}
\put(0.79,0.233){\large $\tau = 1$}
\end{picture}
\caption{Causal $3$-simplices employed in $(2+1)$-dimensional causal dynamical triangulations extending from time slice $\tau=0$ to time slice $\tau=1$. (a) $(3,1)$ $3$-simplex, (b) $(2,2)$ $3$-simplex, (c) $(1,3)$ $3$-simplex. We have adapted this figure from \cite{JHC&JMM}.}
\label{3-simplices}
\end{figure}

Within the causal dynamical triangulations approach to such a quantization,\footnote{See \cite{JA&JJ&RL1,JA&JJ&RL2,JA&RL} for the original formulation and \cite{JA&AG&JJ&RL3} for a comprehensive review.} one restricts the path integration in equation \eqref{gravitypathintegral} to so-called causal spacetime metric tensors $\mathbf{g}_{c}$, those admitting a global foliation by spacelike hypersurfaces all of a fixed spatial topology $\Sigma$. The manifold $\mathcal{M}$ therefore has the topology $\Sigma\times\mathsf{I}$, the direct product of $\Sigma$ and a real interval $\mathsf{I}$. By invoking this restriction, one considers transition amplitudes $\mathscr{A}_{\Sigma}[\gamma]$ formally defined as 
\begin{equation}\label{causalgravitypathintegral}
\mathscr{A}_{\Sigma}[\gamma]=\int_{\substack{\mathcal{M}\cong\Sigma\times\mathsf{I} \\ \mathbf{g}_{c}|_{\partial\mathcal{M}}=\gamma}}\mathrm{d}\mu(\mathbf{g}_{c})\,e^{iS_{\mathrm{cl}}[\mathbf{g}_{c}]/\hbar}.
\end{equation}
To regularize the transition amplitudes $\mathscr{A}_{\Sigma}[\gamma]$, 
one replaces the path integration over all causal metric tensors $\mathbf{g}_{c}$ in equation \eqref{causalgravitypathintegral} with a path summation over all causal triangulations $\mathcal{T}_{c}$. A causal triangulation $\mathcal{T}_{c}$ is a piecewise-Minkowski simplicial manifold possessing a global foliation by spacelike hypersurfaces all of the topology $\Sigma$. One constructs a causal triangulation $\mathcal{T}_{c}$ by appropriately gluing together $N_{D}$ causal $D$-simplices, each a simplicial piece of $D$-dimensional Minkowski spacetime with spacelike edges of squared invariant length $a^{2}$ and timelike edges of squared invariant length $-\alpha a^{2}$ for positive constant $\alpha$. $a$ is the lattice spacing. We depict the three types of causal $3$-simplices (tetrahedra) in figure \ref{3-simplices}. 
Causal $D$-simplices necessarily assemble into a manifold of topology $\Sigma\times\mathsf{I}$, and their skeleton distinguishes a foliation of this manifold into spacelike hypersurfaces. We refer to the leaves of this distinguished foliation as a causal triangulation's time slices, and we enumerate a causal triangulation's $T$ time slices with a discrete time coordinate $\tau$.

By invoking this regularization, one considers regularized transition amplitudes $\mathcal{A}_{\Sigma}[\Gamma]$ defined as
\begin{equation}\label{causalpathsum}
\mathcal{A}_{\Sigma}[\Gamma]=\sum_{\substack{\mathcal{T}_{c}\cong\Sigma\times\mathsf{I} \\ \mathcal{T}_{c}|_{\partial\mathcal{T}_{c}}=\Gamma}}\mu(\mathcal{T}_{c})\,e^{i\mathcal{S}_{\mathrm{cl}}[\mathcal{T}_{c}]/\hbar}.
\end{equation} 
$\Gamma$ is the triangulation of the boundary $\partial\mathcal{T}_{c}$ of the causal triangulation $\mathcal{T}_{c}$, $\mu(\mathcal{T}_{c})$ is the measure, equal to the inverse of the order of the automorphism group of the causal triangulation $\mathcal{T}_{c}$, and $\mathcal{S}_{\mathrm{cl}}[\mathcal{T}_{c}]$ is the translation of the action $S_{\mathrm{cl}}[\mathbf{g}]$ into the Regge calculus of causal triangulations. In the cases of $D>2$ dimensions, analytic calculations of the transition amplitudes $\mathcal{A}_{\Sigma}[\Gamma]$, 
even for the simplest nontrivial cases, are not currently possible. To study the quantum theory of gravity defined by the transition amplitudes $\mathcal{A}_{\Sigma}[\Gamma]$, 
one therefore employs numerical techniques, specifically Monte Carlo methods. To enable the application of such methods, one first performs a Wick rotation of each causal triangulation by analytically continuing $\alpha$ to $-\alpha$ through the lower-half complex plane. This Wick rotation transforms the transition amplitude $\mathcal{A}_{\Sigma}[\Gamma]$ 
into the partition function
\begin{equation}\label{partitionfunction}
\mathcal{Z}_{\Sigma}[\Gamma]=\sum_{\substack{\mathcal{T}_{c}\cong\Sigma\times\mathsf{I} \\ \mathcal{T}_{c}|_{\partial\mathcal{T}_{c}}=\Gamma}}\mu(\mathcal{T}_{c})\,e^{-\mathcal{S}_{\mathrm{cl}}^{(\mathrm{E})}[\mathcal{T}_{c}]}
\end{equation}
in which $\mathcal{S}_{\mathrm{cl}}^{(\mathrm{E})}[\mathcal{T}_{c}]$ is the resulting real-valued Euclidean action. Since one can only numerically simulate finite causal triangulations, one chooses to consider the partition function \eqref{partitionfunction} for fixed numbers $\bar{T}$ of time slices and $\bar{N}_{D}$ of causal $D$-simplices. Accordingly, Monte Carlo methods produce ensembles of causal triangulations representative of those contributing to the (canonical) partition function
\begin{equation}\label{partitionfunctionfixedTN}
Z_{\Sigma}[\Gamma]=\sum_{\substack{\mathcal{T}_{c}\cong\Sigma\times\mathsf{I} \\ \mathcal{T}_{c}|_{\partial\mathcal{T}_{c}}=\Gamma \\ T(\mathcal{T}_{c})=\bar{T} \\ N_{D}(\mathcal{T}_{c})=\bar{N}_{D}}}\mu(\mathcal{T}_{c})\,e^{-\mathcal{S}_{\mathrm{cl}}^{(\mathrm{E})}[\mathcal{T}_{c}]},
\end{equation}
related by Laplace transform to the (grand canonical) partition function \eqref{partitionfunction}. 

We take the action $S_{\mathrm{cl}}[\mathbf{g}]$ as that of $(2+1)$-dimensional Einstein gravity:
\begin{equation}\label{completeCaction}
S_{\mathrm{cl}}[\mathbf{g}]=\frac{1}{16\pi G_{0}}\int_{\mathcal{M}}\mathrm{d}^{3}x\,\sqrt{-g}\left(R-2\Lambda_{0}\right)+\frac{1}{8\pi G_{0}}\int_{\partial\mathcal{M}}\mathrm{d}^{2}y\sqrt{|\gamma|}K.
\end{equation}
The first term in the action \eqref{completeCaction}---the bulk term---is the Einstein-Hilbert action in which $G_{0}$ is the bare Newton constant, $R$ is the Ricci scalar of the metric tensor $\mathbf{g}$, and $\Lambda_{0}$ is a positive bare cosmological constant.  The second term in the action \eqref{completeCaction}---the boundary term---is the Gibbons-Hawking-York action in which $K$ is the trace of the extrinsic curvature of the metric tensor $\gamma$ \cite{GWG&SWH,JWY}. We choose to consider a spacetime manifold $\mathcal{M}$ isomorphic to the direct product $\mathsf{S}^{2}\times\mathsf{I}$ of a $2$-sphere $\mathsf{S}^{2}$ and a real interval $\mathsf{I}$. In this case the boundary $\partial\mathcal{T}_{c}$ consists of two disconnected components: an initial spacelike $2$-sphere $\mathsf{S}_{\mathrm{i}}^{2}$ and a final spacelike $2$-sphere $\mathsf{S}_{\mathrm{f}}^{2}$. Drawing on previous results of Hartle and Sorkin \cite{JBH&RS}, Ambj\o rn \emph{et al} \cite{JA&JJ&RL2}, and Anderson \emph{et al} \cite{CA&SJC&JHC&PH&RKK&PZ}, Cooperman and Miller derived the form of the action $\mathcal{S}_{\mathrm{cl}}^{(\mathrm{E})}[\mathcal{T}_{c}]$ arising from the action \eqref{completeCaction} for this case. We display $\mathcal{S}_{\mathrm{cl}}^{(\mathrm{E})}[\mathcal{T}_{c}]$ in equation \eqref{completeEaction} of appendix \ref{completediscreteEaction}. If the initial and final boundary $2$-spheres $\mathsf{S}_{\mathrm{i}}^{2}$ and $\mathsf{S}_{\mathrm{f}}^{2}$ are identified, yielding periodic boundary conditions in the temporal direction, then the action $\mathcal{S}_{\mathrm{cl}}^{(\mathrm{E})}[\mathcal{T}_{c}]$ simplifies considerably \cite{JA&JJ&RL2}: 
\begin{equation}\label{CDTaction3}
\mathcal{S}_{\mathrm{cl}}^{(\mathrm{E})}[\mathcal{T}_{c}]=-k_{0}N_{0}+k_{3}N_{3}.
\end{equation}
$N_{0}$ is the number of $0$-simplices (vertices), $N_{3}$ is the number of $3$-simplices, and the bare couplings $k_{0}$ and $k_{3}$ are the following dimensionless combinations of $G_{0}$, $\Lambda_{0}$, and $a$:
\begin{subequations}\label{k0k3expressions}
\begin{eqnarray}
k_{0}&=&2\pi ak\\
k_{3}&=&\frac{a^{3}\lambda}{4\sqrt{2}}+2\pi ak\left[\frac{3}{\pi}\cos^{-1}{\left(\frac{1}{3}\right)}-1\right]
\end{eqnarray}
\end{subequations}
with 
\begin{subequations}
\begin{eqnarray}
k&=&\frac{1}{8\pi G_{0}}\\ 
\lambda&=&\frac{\Lambda_{0}}{8\pi G_{0}}
\end{eqnarray}
\end{subequations}
We set $\alpha=1$ because the value of $\alpha$ (once the Wick rotation has been performed) is irrelevant in $2+1$ dimensions. 
When referring to an ensemble of causal triangulations with fixed initial and final boundary $2$-spheres $\mathsf{S}_{\mathrm{i}}^{2}$ and $\mathsf{S}_{\mathrm{f}}^{2}$, we employ the couplings $k_{0}$ and $k_{3}$ instead of the couplings $k$ and $\lambda$ of equation \eqref{completeEaction} to facilitate contact with previous work. By the given values of $k_{0}$ and $k_{3}$, we mean the values dictated by the relations \eqref{k0k3expressions} for the values of $k$ and $\lambda$ actually characterizing the given ensemble. An ensemble of causal triangulations 
is therefore characterized by the number $\bar{T}$ of time slices, the number $\bar{N}_{3}$ of $3$-simplices, the value of the coupling $k_{0}$, and the triangulations $\Gamma(\mathsf{S}_{\mathrm{i}}^{2})$ and $\Gamma(\mathsf{S}_{\mathrm{f}}^{2})$ of the initial and final boundary $2$-spheres $\mathsf{S}_{\mathrm{i}}^{2}$ and $\mathsf{S}_{\mathrm{f}}^{2}$. As explained, for instance in \cite{JHC&JMM}, we must tune the coupling $k_{3}$ to its critical value $k_{3}^{c}$ to ensure that the partition function \eqref{partitionfunctionfixedTN} for the action \eqref{completeEaction} is well-defined. The value $k_{3}^{c}$ is therefore not independent of the other quantities characterizing an ensemble of causal triangulations.



The triangulations $\Gamma(\mathsf{S}_{\mathrm{i}}^{2})$ and $\Gamma(\mathsf{S}_{\mathrm{f}}^{2})$ completely characterize the geometries of the initial and final boundary $2$-spheres $\mathsf{S}_{\mathrm{i}}^{2}$ and $\mathsf{S}_{\mathrm{f}}^{2}$, constituting a sizeable amount of boundary data on which the partition function \eqref{partitionfunctionfixedTN} depends. Cooperman and Miller restricted attention to only one aspect of the geometries of the triangulations $\Gamma(\mathsf{S}_{\mathrm{i}}^{2})$ and $\Gamma(\mathsf{S}_{\mathrm{f}}^{2})$: their discrete spatial $2$-volumes as measured by the numbers $N_{2}^{\mathrm{SL}}(\mathsf{S}_{\mathrm{i}}^{2})$ and $N_{2}^{\mathrm{SL}}(\mathsf{S}_{\mathrm{f}}^{2})$ of spacelike $2$-simplices (equilateral triangles) comprising the $2$-spheres $\mathsf{S}_{\mathrm{i}}^{2}$ and $\mathsf{S}_{\mathrm{f}}^{2}$. The dependence of the partition function \eqref{partitionfunctionfixedTN} on $N_{2}^{\mathrm{SL}}(\mathsf{S}_{\mathrm{i}}^{2})$ and $N_{2}^{\mathrm{SL}}(\mathsf{S}_{\mathrm{f}}^{2})$ is not merely the simplest to consider: in the absence of a physically relevant characterization of the geometries of the triangulations $\Gamma(\mathsf{S}_{\mathrm{i}}^{2})$ and $\Gamma(\mathsf{S}_{\mathrm{f}}^{2})$, the dependence of the partition function \eqref{partitionfunctionfixedTN} on other aspects of these geometries is difficult to study meaningfully. 
To probe only the dependence on the initial and final numbers $N_{2}^{\mathrm{SL}}(\mathsf{S}_{\mathrm{i}}^{2})$ and $N_{2}^{\mathrm{SL}}(\mathsf{S}_{\mathrm{f}}^{2})$ of spacelike $2$-simplices, Cooperman and Miller proceeded as follows. They generated $\mathsf{N}$ random triangulations $\Gamma(\mathsf{S}_{\mathrm{i}}^{2})$ of the $2$-sphere $\mathsf{S}_{\mathrm{i}}^{2}$ constructed from precisely $N_{2}^{\mathrm{SL}}(\mathsf{S}_{\mathrm{i}}^{2})$ spacelike $2$-simplices and $\mathsf{N}$ random triangulations $\Gamma(\mathsf{S}_{\mathrm{f}}^{2})$ of the $2$-sphere $\mathsf{S}_{\mathrm{f}}^{2}$ constructed from precisely $N_{2}^{\mathrm{SL}}(\mathsf{S}_{\mathrm{f}}^{2})$ spacelike $2$-simplices; they randomly paired the former $\mathsf{N}$ triangulations with the latter $\mathsf{N}$ triangulations to form $\mathsf{N}$ pairs of initial and final boundary triangulations $\Gamma(\mathsf{S}_{\mathrm{i}}^{2})$ and $\Gamma(\mathsf{S}_{\mathrm{f}}^{2})$; they generated an ensemble of causal triangulations for each of these $\mathsf{N}$ pairs; and 
they combined these $\mathsf{N}$ ensembles into a single averaged ensemble.
\footnote{Technically, the procedure of Cooperman and Miller assumes a constant measure over all causal triangulations with initial and final boundary triangulations $\Gamma(\mathsf{S}_{\mathrm{i}}^{2})$ and $\Gamma(\mathsf{S}_{\mathrm{f}}^{2})$ constructed respectively from precisely $N_{2}^{\mathrm{SL}}(\mathsf{S}_{\mathrm{i}}^{2})$ and $N_{2}^{\mathrm{SL}}(\mathsf{S}_{\mathrm{f}}^{2})$ spacelike $2$-simplices (for given values of $\bar{T}$, $\bar{N}_{3}$, and $k_{0}$)  \cite{JHC&JMM}.} 

By choosing to consider the dependence of the partition function \eqref{partitionfunctionfixedTN} only on $N_{2}^{\mathrm{SL}}(\mathsf{S}_{\mathrm{i}}^{2})$ and $N_{2}^{\mathrm{SL}}(\mathsf{S}_{\mathrm{f}}^{2})$, Cooperman and Miller emulated virtually all previous studies 
of causal dynamical triangulations in $2+1$ dimensions (and in $3+1$ dimensions) in probing the ground state of the quantum geometry defined by an ensembles of causal triangulations. Prior investigations examined the spacetime manifold structure $\mathrm{S}^{2}\times\mathrm{S}^{1}$ for which 
the temporal direction is periodically identified. Such studies probe the ground state of quantum geometry in the sense that there are no boundary conditions to induce excitations of the quantum geometry. Although Cooperman and Miller explored transition amplitudes with the spacetime manifold structure $\mathsf{S}^{2}\times\mathsf{I}$ in \cite{JHC&JMM}, their averaging 
over all geometrical degrees of freedom of the boundary $2$-spheres except for their discrete spatial $2$-volumes results in boundary conditions that do not induce excitations of the quantum geometry.

Monte Carlo methods do not give us access to the partition function \eqref{partitionfunctionfixedTN} itself; they yield only a representative sample of causal triangulations contributing to the path summation defining the partition function \eqref{partitionfunctionfixedTN}. This fact poses no problem of principle: we do have access to the expectation values of observables in the quantum state defined by the partition function \eqref{partitionfunctionfixedTN}. One computes the expectation value $\mathbb{E}[\mathcal{O}]$ of an observable $\mathcal{O}$ in this quantum state 
as follows:
\begin{equation}
\mathbb{E}[\mathcal{O}]=\frac{1}{Z_{\Sigma}[\Gamma]}\sum_{\substack{\mathcal{T}_{c}\cong\Sigma\times\mathsf{I} \\ \mathcal{T}_{c}|_{\partial\mathcal{T}_{c}}=\Gamma \\ T(\mathcal{T}_{c})=\bar{T} \\ N_{D}(\mathcal{T}_{c})=\bar{N}_{D}}}\mu(\mathcal{T}_{c})\,e^{-\mathcal{S}_{\mathrm{cl}}^{(\mathrm{E})}[\mathcal{T}_{c}]}\mathcal{O}[\mathcal{T}_{c}].
\end{equation}
We approximate the expectation value $\mathbb{E}[\mathcal{O}]$ by its average
\begin{equation}
\langle\mathcal{O}\rangle=\frac{1}{N(\mathcal{T}_{c})}\sum_{j=1}^{N(\mathcal{T}_{c})}\mathcal{O}[\mathcal{T}_{c}^{(j)}]
\end{equation}
over an ensemble of $N(\mathcal{T}_{c})$ causal triangulations generated by Monte Carlo methods. The Metropolis algorithm behind these simulations guarantees that 
\begin{equation}
\mathbb{E}[\mathcal{O}]=\lim_{N(\mathcal{T}_{c})\rightarrow\infty}\langle\mathcal{O}\rangle.
\end{equation}
Numerical measurements of certain observables' ensemble averages have revealed that the model defined by the partition function \eqref{partitionfunctionfixedTN} for the action \eqref{CDTaction3} exhibits two phases of quantum geometry separated by a first-order phase transition: the decoupled phase, labeled A, for coupling $k_{0}>k_{0}^{c}$ and the condensate phase, labeled C, for coupling $k_{0}<k_{0}^{c}$ \cite{JA&JJ&RL3,RK}. Cooperman and Miller found that phase C also exists within the model defined by the partition function \eqref{partitionfunctionfixedTN} for the action \eqref{completeEaction} \cite{JHC&JMM}. 
We restrict attention to values of the coupling $k_{0}$ that fall within phase C as only the quantum geometry defined by ensembles of causal triangulations with phase C possesses physical properties. We explore these properties in sections \ref{evidenceconjecture}, \ref{analysissupport}, and \ref{argumentrefutation}.

\section{Evidence and conjecture}\label{evidenceconjecture}

We now review and expand upon the evidence that led Cooperman and Miller to formulate their conjecture. Following several previous authors \cite{JA&JGS&AG&JJ,JA&JGS&AG&JJ2,JA&AG&JJ&RL1,JA&AG&JJ&RL2,JA&AG&JJ&RL&JGS&TT,JA&JJ&RL3,JA&JJ&RL4,JA&JJ&RL5,JA&JJ&RL6,CA&SJC&JHC&PH&RKK&PZ,RK}, Cooperman and Miller performed measurements of the number $N_{2}^{\mathrm{SL}}(\tau)$ of spacelike $2$-simplices as a function of the discrete time coordinate $\tau$ labeling the distinguished foliation's time slices 
\cite{JHC&JMM}. $N_{2}^{\mathrm{SL}}(\tau)$ quantifies the evolution of discrete spatial $2$-volume in the distinguished foliation. 

\begin{figure}[!ht]
\centering
\setlength{\unitlength}{\textwidth}
\begin{picture}(1,0.2)
\put(0.1,0.005){\includegraphics[scale=1]{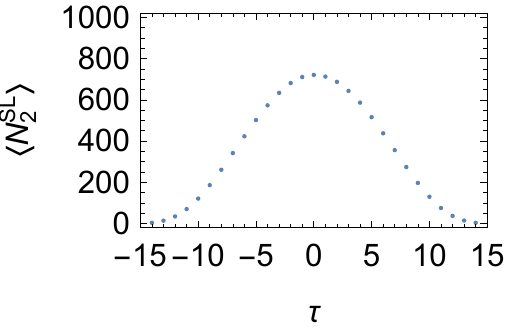}}
\put(0.5,0.005){\includegraphics[scale=1]{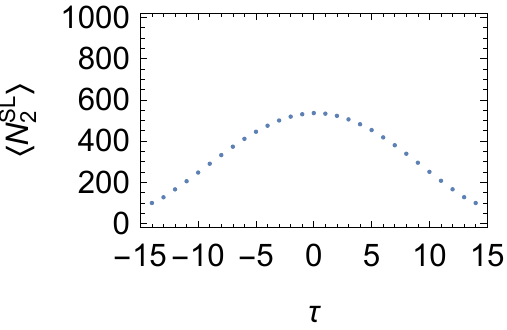}}
\put(0.28,-0.02){(a)}
\put(0.68,-0.02){(b)}
\end{picture}
\caption{Ensemble average number $\langle N_{2}^{\mathrm{SL}}\rangle$ of spacelike $2$-simplices as a function of the discrete time coordinate $\tau$ for $\bar{T}=29$, $\bar{N}_{3}=30850$, and $k_{0}=1.00$. (a) $N_{2}^{\mathrm{SL}}(\mathsf{S}_{\mathrm{i}}^{2})=N_{2}^{\mathrm{SL}}(\mathsf{S}_{\mathrm{f}}^{2})=4$  (b) $N_{2}^{\mathrm{SL}}(\mathsf{S}_{\mathrm{i}}^{2})=N_{2}^{\mathrm{SL}}(\mathsf{S}_{\mathrm{f}}^{2})=100$. We have taken this data from \cite{JHC&JMM}.}
\label{nonminnonminsame1}
\end{figure}

Cooperman and Miller first considered the following two ensembles of causal triangulations. For $\bar{T}=29$, $\bar{N}_{3}=30850$, and $k_{0}=1.00$, we display the ensemble average $\langle N_{2}^{\mathrm{SL}}(\tau)\rangle$ for $N_{2}^{\mathrm{SL}}(\mathsf{S}_{\mathrm{i}}^{2})=N_{2}^{\mathrm{SL}}(\mathsf{S}_{\mathrm{f}}^{2})=4$\footnote{The minimal piecewise-Euclidean simplicial $2$-sphere is constructed from four $2$-simplices.} in figure \ref{nonminnonminsame1}(a) and for $N_{2}^{\mathrm{SL}}(\mathsf{S}_{\mathrm{i}}^{2})=N_{2}^{\mathrm{SL}}(\mathsf{S}_{\mathrm{f}}^{2})=100$ in figure \ref{nonminnonminsame1}(b).
The plot in figure \ref{nonminnonminsame1}(a) shows the behavior of $\langle N_{2}^{\mathrm{SL}}(\tau)\rangle$ previously understood as characteristic of phase C \cite{JA&AG&JJ&AK&RL,JA&AG&JJ&RL1,JA&AG&JJ&RL2,JA&AG&JJ&RL&JGS&TT,JA&JJ&RL3,JA&JJ&RL4,JA&JJ&RL5,JA&JJ&RL6,CA&SJC&JHC&PH&RKK&PZ,DB&JH2,JHC,JHC&JMM,RK}: $\langle N_{2}^{\mathrm{SL}}(\tau)\rangle$ smoothly increases from its minimal value of $4$ at the initial boundary $2$-sphere $\mathsf{S}_{\mathrm{i}}^{2}$ to its maximal value at the central time slice and symmetrically decreases from its maximal value to its minimal value of $4$ at the final boundary $2$-sphere $\mathsf{S}_{\mathrm{f}}^{2}$. As several authors have previously demonstrated \cite{JA&AG&JJ&AK&RL,JA&AG&JJ&RL1,JA&AG&JJ&RL2,JA&JJ&RL3,JA&JJ&RL4,JA&JJ&RL5,JA&JJ&RL6,CA&SJC&JHC&PH&RKK&PZ,DB&JH2,JHC,JHC&JMM,RK}, and as we demonstrate once more in section \ref{analysissupport}, the ground state solution---Euclidean de Sitter space---of a minisuperspace model based on Euclidean Einstein gravity accurately describes the shape of $\langle N_{2}^{\mathrm{SL}}(\tau)\rangle$. The plot in figure \ref{nonminnonminsame1}(b) shows that the characteristic behavior of $\langle N_{2}^{\mathrm{SL}}(\tau)\rangle$ continues to be manifest even for boundary $2$-spheres with nonminimal discrete spatial $2$-volumes. Cooperman and Miller demonstrated, moreover, that a portion of Euclidean de Sitter space accurately describes the shape of $\langle N_{2}^{\mathrm{SL}}(\tau)\rangle$ in this case as well \cite{JHC&JMM}. 

\begin{figure}[!ht]
\centering
\setlength{\unitlength}{\textwidth}
\begin{picture}(1,0.22)
\put(0,0.005){\includegraphics[scale=1]{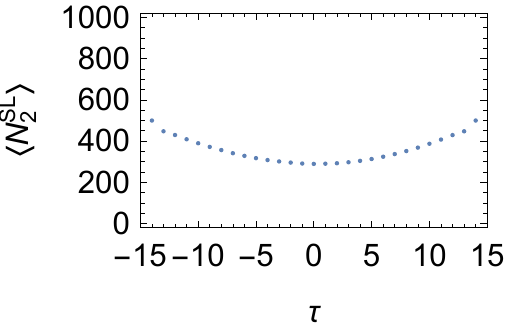}}
\put(0.34,0.005){\includegraphics[scale=1]{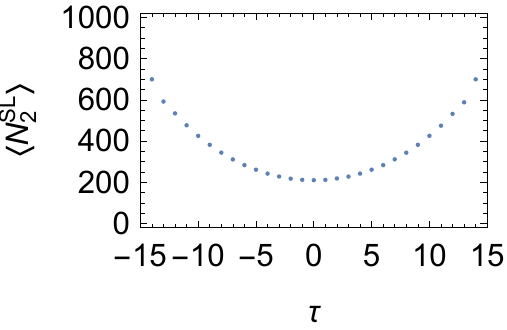}}
\put(0.67,0.005){\includegraphics[scale=1]{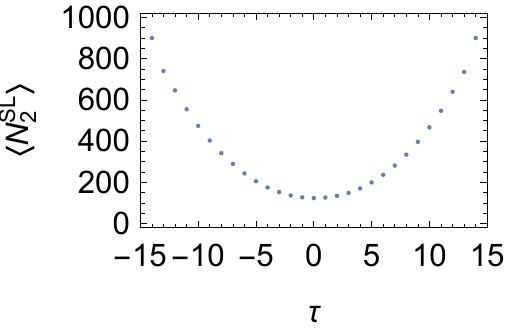}}
\put(0.18,-0.02){(a)}
\put(0.52,-0.02){(b)}
\put(0.85,-0.02){(c)}
\end{picture}
\caption{Ensemble average number $\langle N_{2}^{\mathrm{SL}}\rangle$ of spacelike $2$-simplices as a function of the discrete time coordinate $\tau$ for $\bar{T}=29$, $\bar{N}_{3}=30850$, and $k_{0}=1.00$. (a) $N_{2}^{\mathrm{SL}}(\mathsf{S}_{\mathrm{i}}^{2})=N_{2}^{\mathrm{SL}}(\mathsf{S}_{\mathrm{f}}^{2})=500$  (b) $N_{2}^{\mathrm{SL}}(\mathsf{S}_{\mathrm{i}}^{2})=N_{2}^{\mathrm{SL}}(\mathsf{S}_{\mathrm{f}}^{2})=700$ (c) $N_{2}^{\mathrm{SL}}(\mathsf{S}_{\mathrm{i}}^{2})=N_{2}^{\mathrm{SL}}(\mathsf{S}_{\mathrm{f}}^{2})=900$. We have taken this data from \cite{JHC&JMM}.}
\label{nonminnonminsame2}
\end{figure}

\begin{figure}[!ht]
\centering
\setlength{\unitlength}{\textwidth}
\begin{picture}(1,0.2)
\put(0.1,0.005){\includegraphics[scale=1]{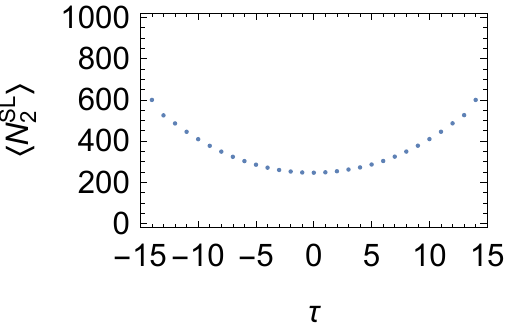}}
\put(0.5,0.005){\includegraphics[scale=1]{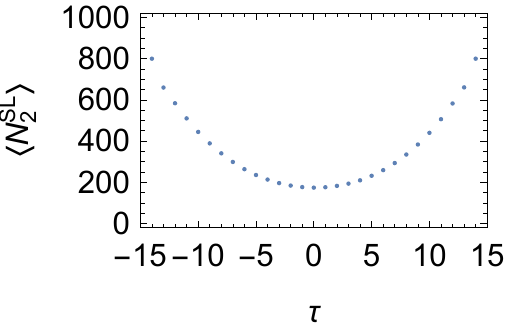}}
\put(0.28,-0.02){(a)}
\put(0.68,-0.02){(b)} 
\end{picture}
\caption{Ensemble average number $\langle N_{2}^{\mathrm{SL}}\rangle$ of spacelike $2$-simplices as a function of the discrete time coordinate $\tau$ for $\bar{T}=29$, $\bar{N}_{3}=30850$, and $k_{0}=1.00$. (a) $N_{2}^{\mathrm{SL}}(\mathsf{S}_{\mathrm{i}}^{2})=N_{2}^{\mathrm{SL}}(\mathsf{S}_{\mathrm{f}}^{2})=600$ (b) $N_{2}^{\mathrm{SL}}(\mathsf{S}_{\mathrm{i}}^{2})=N_{2}^{\mathrm{SL}}(\mathsf{S}_{\mathrm{f}}^{2})=800$.}
\label{nonminnonmin2}
\end{figure}

Cooperman and Miller next increased further the discrete spatial $2$-volumes of the initial and final boundary $2$-spheres. For $\bar{T}=29$, $\bar{N}_{3}=30850$, $k_{0}=1.00$, we display $\langle N_{2}^{\mathrm{SL}}(\tau)\rangle$ for $N_{2}^{\mathrm{SL}}(\mathsf{S}_{\mathrm{i}}^{2})=N_{2}^{\mathrm{SL}}(\mathsf{S}_{\mathrm{f}}^{2})=500$ in figure \ref{nonminnonminsame2}(a), for $N_{2}^{\mathrm{SL}}(\mathsf{S}_{\mathrm{i}}^{2})=N_{2}^{\mathrm{SL}}(\mathsf{S}_{\mathrm{f}}^{2})=700$ in figure \ref{nonminnonminsame2}(b), and for $N_{2}^{\mathrm{SL}}(\mathsf{S}_{\mathrm{i}}^{2})=N_{2}^{\mathrm{SL}}(\mathsf{S}_{\mathrm{f}}^{2})=900$ in figure \ref{nonminnonminsame2}(c).\footnote{The ensemble of causal triangulations characterized by $\bar{T}=29$, $\bar{N}_{3}=30850$, $k_{0}=1.00$, and $N_{2}^{\mathrm{SL}}(\mathsf{S}_{\mathrm{i}}^{2})=N_{2}^{\mathrm{SL}}(\mathsf{S}_{\mathrm{f}}^{2})=300$ is very close to the transition of $\langle N_{2}^{\mathrm{SL}}(\tau)\rangle$ from being concave-down to being concave-up for these values of $\bar{T}$, $\bar{N}_{3}$ and $k_{0}$. We have not yet performed Monte Carlo simulations for sufficiently long computer times to determine on which side of the transition this ensemble falls.}
We considered two further ensembles of causal triangulations. 
For $\bar{T}=29$, $\bar{N}_{3}=30850$, and $k_{0}=1.00$, we display $\langle N_{2}^{\mathrm{SL}}(\tau)\rangle$ for $N_{2}^{\mathrm{SL}}(\mathsf{S}_{\mathrm{i}}^{2})=N_{2}^{\mathrm{SL}}(\mathsf{S}_{\mathrm{f}}^{2})=600$ in figure \ref{nonminnonmin2}(a) and for $N_{2}^{\mathrm{SL}}(\mathsf{S}_{\mathrm{i}}^{2})=N_{2}^{\mathrm{SL}}(\mathsf{S}_{\mathrm{f}}^{2})=800$ in figure \ref{nonminnonmin2}(b). 
As Cooperman and Miller remarked, the shape of $\langle N_{2}^{\mathrm{SL}}(\tau)\rangle$ for those ensembles represented in figures \ref{nonminnonminsame2} 
and \ref{nonminnonmin2} is possibly of a hyperbolic sinusoidal character. They hypothesized accordingly that a portion of Lorentzian de Sitter spacetime might accurately describe the shape of $\langle N_{2}^{\mathrm{SL}}(\tau)\rangle$ for these ensembles \cite{JHC&JMM}. We test this hypothesis in section \ref{analysissupport}. 

\begin{figure}[!ht]
\centering
\setlength{\unitlength}{\textwidth}
\begin{picture}(1,0.5)
\put(0,0.25){\includegraphics[scale=0.9]{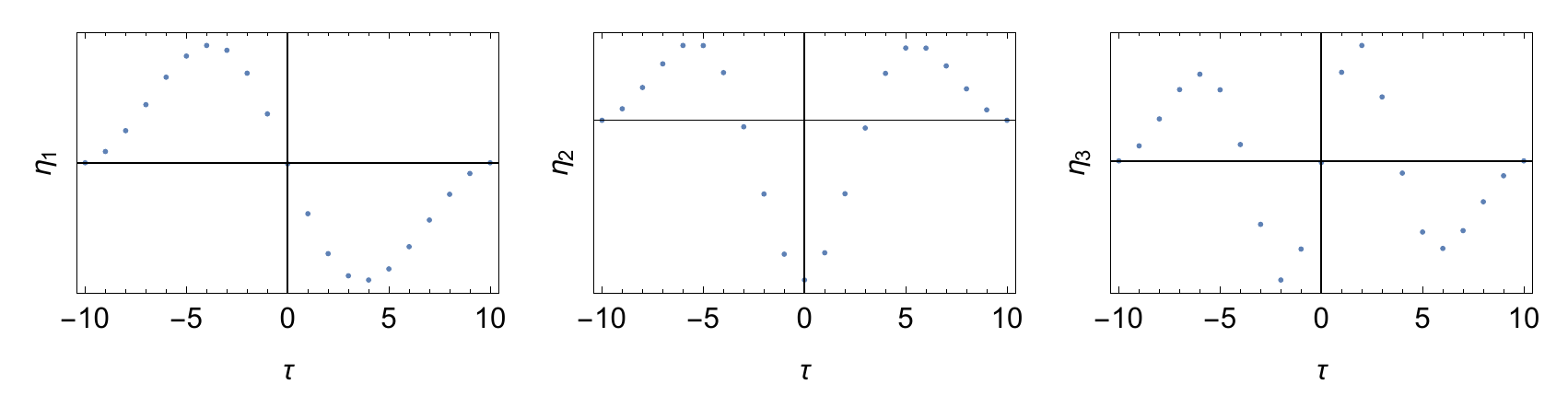}}
\put(0,-0.02){\includegraphics[scale=0.9]{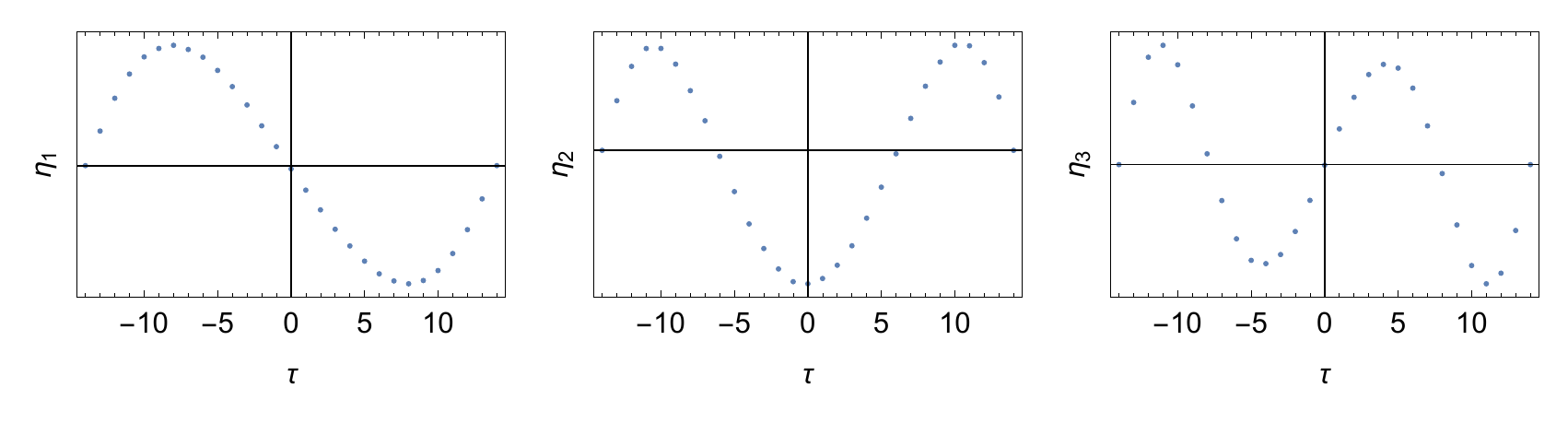}}
\put(0.47,-0.035){(b)}
\put(0.47,0.235){(a)}
\end{picture}
\caption{First three eigenvectors $\eta_{j}(\tau)$ of the ensemble average connected $2$-point function $\langle n_{2}^{\mathrm{SL}}(\tau)\,n_{2}^{\mathrm{SL}}(\tau')\rangle$ of deviations $n_{2}^{\mathrm{SL}}$ in the number of spacelike $2$-simplices as a function of the discrete time coordinate $\tau$ for $\bar{N}_{3}=30850$ and $k_{0}=1.00$ (a) $\bar{T}=21$ and $N_{2}^{\mathrm{SL}}(\mathsf{S}_{\mathrm{i}}^{2})=N_{2}^{\mathrm{SL}}(\mathsf{S}_{\mathrm{f}}^{2})=4$ (ensemble $\mathcal{E}_{\mathrm{E}}$) (b) $\bar{T}=29$ and $N_{2}^{\mathrm{SL}}(\mathsf{S}_{\mathrm{i}}^{2})=N_{2}^{\mathrm{SL}}(\mathsf{S}_{\mathrm{f}}^{2})=600$ (ensemble $\mathcal{E}_{\mathrm{L}}$). We do not indicate the scale of the eigenvectors $\eta_{j}(\tau)$ as their normalization is arbitrary.}
\label{eigenvectors}
\end{figure}

\begin{figure}[!ht]
\setlength{\unitlength}{\textwidth}
\begin{picture}(1,0.3)
\put(0.0005,0.005){\includegraphics[scale=1]{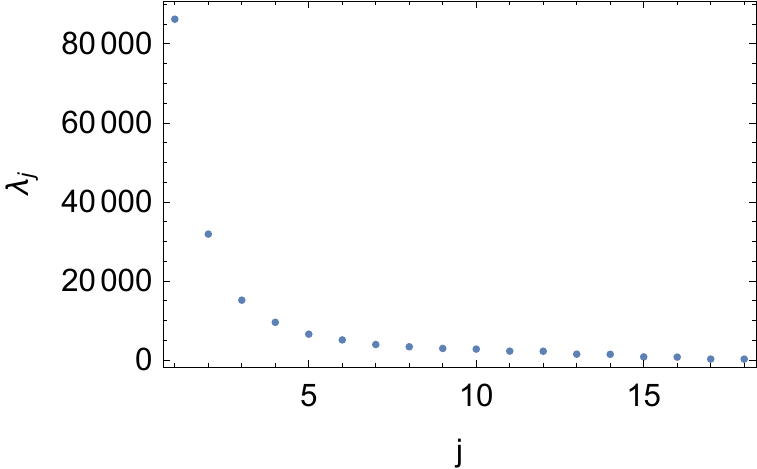}}
\put(0.51,0.00){\includegraphics[scale=1]{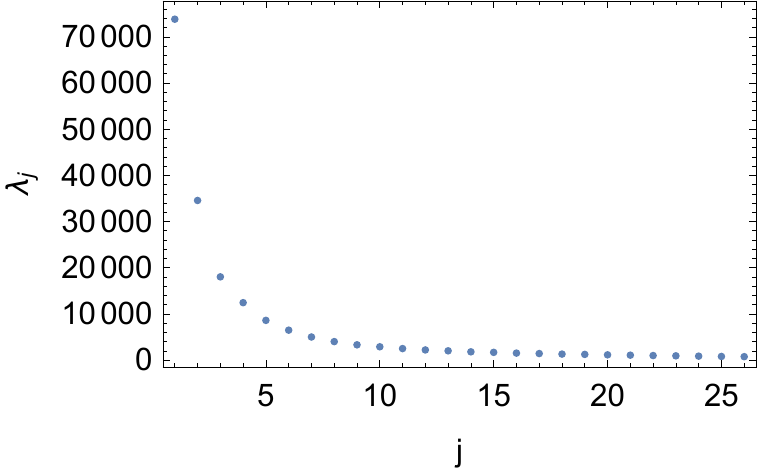}}
\put(0.27,-0.03){(a)}
\put(0.78,-0.03){(b)}
\end{picture}
\caption{Eigenvalues $\lambda_{j}$ of the ensemble average connected $2$-point function $\langle n_{2}^{\mathrm{SL}}(\tau)\,n_{2}^{\mathrm{SL}}(\tau')\rangle$ of deviations $n_{2}^{\mathrm{SL}}$ in the number of spacelike $2$-simplices as a function of the discrete time coordinate $\tau$ for $\bar{N}_{3}=30850$ and $k_{0}=1.00$ (a) $\bar{T}=21$ and $N_{2}^{\mathrm{SL}}(\mathsf{S}_{\mathrm{i}}^{2})=N_{2}^{\mathrm{SL}}(\mathsf{S}_{\mathrm{f}}^{2})=4$ (ensemble $\mathcal{E}_{\mathrm{E}}$) (b) $\bar{T}=29$ and $N_{2}^{\mathrm{SL}}(\mathsf{S}_{\mathrm{i}}^{2})=N_{2}^{\mathrm{SL}}(\mathsf{S}_{\mathrm{f}}^{2})=600$ (ensemble $\mathcal{E}_{\mathrm{L}}$).}
\label{eigenvalues}
\end{figure}

Following Ambj\o rn \emph{et al} \cite{JA&AG&JJ&RL1,JA&AG&JJ&RL2} and Cooperman \cite{JHC}, we moreover measured the ensemble average connected $2$-point function $\langle n_{2}^{\mathrm{SL}}(\tau)\,n_{2}^{\mathrm{SL}}(\tau')\rangle$ of deviations $n_{2}^{\mathrm{SL}}(\tau)$ in the number $N_{2}^{\mathrm{SL}}(\tau)$ of spacelike $2$-simplices from the ensemble average $\langle N_{2}^{\mathrm{SL}}(\tau)\rangle$ defined as
\begin{equation}
\langle n_{2}^{\mathrm{SL}}(\tau)\,n_{2}^{\mathrm{SL}}(\tau')\rangle=\frac{1}{N(\mathcal{T}_{c})}\sum_{j=1}^{N(\mathcal{T}_{c})}\left[n_{2}^{\mathrm{SL}}(\tau)\right]_{j}\left[n_{2}^{\mathrm{SL}}(\tau')\right]_{j}
\end{equation}
for
\begin{equation}
\left[n_{2}^{\mathrm{SL}}(\tau)\right]_{j}=\left[N_{2}^{\mathrm{SL}}(\tau)\right]_{j}-\langle N_{2}^{\mathrm{SL}}(\tau)\rangle.
\end{equation}
$\langle n_{2}^{\mathrm{SL}}(\tau)\,n_{2}^{\mathrm{SL}}(\tau')\rangle$ is a $\bar{T}\times\bar{T}$ real symmetric matrix, which we diagonalize to obtain its eigenvectors $\eta_{j}(\tau)$ and associated eigenvalues $\lambda_{j}$. 
For the (Euclidean-like) ensemble $\mathcal{E}_{\mathrm{E}}$ of causal triangulations characterized by $\bar{T}=21$, $\bar{N}_{3}=30850$, $k_{0}=1.00$, and $N_{2}^{\mathrm{SL}}(\mathsf{S}_{\mathrm{i}}^{2})=N_{2}^{\mathrm{SL}}(\mathsf{S}_{\mathrm{f}}^{2})=4$, we display the first three eigenvectors $\eta_{j}(\tau)$ and the eigenvalues $\lambda_{j}$ of $\langle n_{2}^{\mathrm{SL}}(\tau)\,n_{2}^{\mathrm{SL}}(\tau')\rangle$ in figures \ref{eigenvectors}(a) and \ref{eigenvalues}(a).\footnote{We employ the ensemble $\mathcal{E}_{\mathrm{E}}$ characterized by $\bar{T}=21$, $\bar{N}_{3}=30850$, $k_{0}=1.00$, and $N_{2}^{\mathrm{SL}}(\mathsf{S}_{\mathrm{i}}^{2})=N_{2}^{\mathrm{SL}}(\mathsf{S}_{\mathrm{f}}^{2})=4$ as a point of comparison for two reasons. First, our analysis of $\langle N_{2}^{\mathrm{SL}}(\tau)\rangle$ for the ensemble of causal triangulations characterized by $\bar{T}=29$, $\bar{N}_{3}=30850$, $k_{0}=1.00$, and $N_{2}^{\mathrm{SL}}(\mathsf{S}_{\mathrm{i}}^{2})=N_{2}^{\mathrm{SL}}(\mathsf{S}_{\mathrm{f}}^{2})=4$ indicates the presence of a stalk, resulting in the first eigenvector $\eta_{1}(\tau)$ possessing three rather than two nodes. See \cite{JA&AG&JJ&RL2} for an explanation. Second, our analysis in section \ref{analysissupport} of $\langle N_{2}^{\mathrm{SL}}(\tau)\rangle$ for the ensemble $\mathcal{E}_{\mathrm{E}}$ yields a quality of fit comparable to that for the ensemble $\mathcal{E}_{\mathrm{L}}$ characterized by $\bar{T}=29$, $\bar{N}_{3}=30850$, $k_{0}=1.00$, and $N_{2}^{\mathrm{SL}}(\mathsf{S}_{\mathrm{i}}^{2})=N_{2}^{\mathrm{SL}}(\mathsf{S}_{\mathrm{f}}^{2})=600$, which we anonymously introduced with figure \ref{nonminnonmin2}(a) and formally introduce with figures \ref{eigenvectors}(b) and \ref{eigenvalues}(b).}
The plots in figures \ref{eigenvectors}(a) and \ref{eigenvalues}(a) show the behavior of $\langle n_{2}^{\mathrm{SL}}(\tau)\,n_{2}^{\mathrm{SL}}(\tau')\rangle$ previously understood as characteristic of phase C \cite{JA&AG&JJ&RL1,JA&AG&JJ&RL2,JHC}. As Ambj\o rn \emph{et al} \cite{JA&AG&JJ&RL1,JA&AG&JJ&RL2} and Cooperman \cite{JHC} have previously demonstrated in the case of $3+1$ dimensions, and as we demonstrate for the first time in $2+1$ dimensions in section \ref{analysissupport}, the connected $2$-point function of linear gravitational perturbations propagating on Euclidean de Sitter space accurately describes the shape of $\langle n_{2}^{\mathrm{SL}}(\tau)\,n_{2}^{\mathrm{SL}}(\tau')\rangle$, both its eigenvectors $\eta_{j}(\tau)$ and its eigenvalues $\lambda_{j}$. 

For the (Lorentzian-like) ensemble $\mathcal{E}_{\mathrm{L}}$ of causal triangulations characterized by $\bar{T}=29$, $\bar{N}_{3}=30850$, $k_{0}=1.00$, and $N_{2}^{\mathrm{SL}}(\mathsf{S}_{\mathrm{i}}^{2})=N_{2}^{\mathrm{SL}}(\mathsf{S}_{\mathrm{f}}^{2})=600$, we display the first three eigenvectors $\eta_{j}(\tau)$ and the associated eigenvalues $\lambda_{j}$ of $\langle n_{2}^{\mathrm{SL}}(\tau)\,n_{2}^{\mathrm{SL}}(\tau')\rangle$ in figures \ref{eigenvectors}(b) and \ref{eigenvalues}(b). 
The shapes of the eigenvectors $\eta_{j}(\tau)$ for the ensemble $\mathcal{E}_{\mathrm{L}}$ 
differ subtly yet notably from the shapes of the eigenvectors $\eta_{j}(\tau)$ for the ensemble $\mathcal{E}_{\mathrm{E}}$. 
The spectrum of eigenvalues $\lambda_{j}$ for the ensemble $\mathcal{E}_{\mathrm{L}}$ also differs subtly yet notably from the spectrum of eigenvalues $\lambda_{j}$ for the ensemble $\mathcal{E}_{\mathrm{E}}$. We hypothesize accordingly that linear gravitational perturbations propagating on a portion of Lorentzian de Sitter spacetime might accurately describe the shape of $\langle n_{2}^{\mathrm{SL}}(\tau)\,n_{2}^{\mathrm{SL}}(\tau')\rangle$, both its eigenvectors $\eta_{j}(\tau)$ and its eigenvalues $\lambda_{j}$, for the ensemble $\mathcal{E}_{\mathrm{L}}$. 
We test this hypothesis in section \ref{analysissupport}. 

These finding led Cooperman and Miller to formulate the following conjecture: geometries resembling Lorentzian de Sitter spacetime, not Euclidean de Sitter space, on sufficiently large scales dominate the partition function \eqref{partitionfunction} for the action \eqref{CDTaction3} defining the ground state of $(2+1)$-dimensional causal dynamical triangulations for spherical spatial topology \cite{JHC&JMM}. Cooperman and Miller also suggested that their conjecture's scenario might arise \emph{via} a mechanism similar to that of the Hartle-Hawking no-boundary proposal in which complex geometries contribute to the partition function \cite{JBH&SWH}. 
We subject their conjecture to a first test in section \ref{analysissupport}, obtaining evidence in its favor; however, we argue for a more straightforward explanation of the above findings in section \ref{argumentrefutation}, refuting their conjecture.

\section{Analysis and support}\label{analysissupport}

We now perform a preliminary test of the conjecture of Cooperman and Miller by analyzing the measurements of $\langle N_{2}^{\mathrm{SL}}(\tau)\rangle$ and $\langle n_{2}^{\mathrm{SL}}(\tau)\,n_{2}^{\mathrm{SL}}(\tau')\rangle$ reported in section \ref{evidenceconjecture} on the basis of their conjecture. To connect their conjecture with these measurements, 
we attempt to describe these measurements within a simple yet nontrivial model inspired by their conjecture: a minisuperspace truncation of $(2+1)$-dimensional Einstein gravity having either Lorentzian de Sitter spacetime or Euclidean de Sitter space as its ground state. Several authors have previously employed this model's Euclidean version \cite{JA&DNC&JGS&JJ,JA&JGS&AG&JJ,JA&JGS&AG&JJ2,JA&AG&JJ&AK&RL,JA&AG&JJ&RL1,JA&AG&JJ&RL2,JA&AG&JJ&RL3,JA&AG&JJ&RL&JGS&TT,JA&JJ&RL3,JA&JJ&RL4,JA&JJ&RL5,JA&JJ&RL6,CA&SJC&JHC&PH&RKK&PZ,DB&JH,DB&JH2,JHC,JHC&JMM,RK}, which Ambj\o rn, Jurkiewicz, and Loll first suggested \cite{JA&JJ&RL3,JA&JJ&RL4,JA&JJ&RL5,JA&JJ&RL6}. We specify the model's metric tensor $\mathbf{g}$ by the line element
\begin{equation}\label{minisuperspacemetric}
\mathrm{d}\mathsf{s}^{2}=\pm\omega^{2}\mathrm{d} t^{2}+\rho^{2}( t)\left(\mathrm{d}\theta^{2}+\sin^{2}{\theta}\,\mathrm{d}\phi^{2}\right)
\end{equation} 
for positive constant $\omega$ and scale factor $\rho(t)$ with upper sign ($+$) for Euclidean signature and the lower sign ($-$) for Lorentzian signature. For the line element \eqref{minisuperspacemetric}, expressed in terms of the spatial $2$-volume
\begin{equation}
V_{2}( t)=\int_{0}^{\pi}\mathrm{d}\theta\int_{0}^{2\pi}\mathrm{d}\phi\sqrt{g_{\theta\theta}g_{\phi\phi}}=4\pi \rho^{2}( t),
\end{equation}
the Einstein-Hilbert action, including the Gibbons-Hawking-York action, given in equation \eqref{completeCaction} for Lorentzian signature, becomes
\begin{equation}\label{MSM2action4}
S_{\mathrm{cl}}[V_{2}]=\pm\frac{\omega}{32\pi G}\int_{t_{\mathrm{i}}}^{t_{\mathrm{f}}}\mathrm{d} t\left[\frac{\dot{V}_{2}^{2}(t)}{\omega^{2}V_{2}(t)}\mp4\Lambda V_{2}(t)\right]
\end{equation}
after integration by parts. As in equation \eqref{minisuperspacemetric}, the upper signs correspond to  Euclidean signature, and the lower signs correspond to Lorentzian signature.\footnote{Typically, in Euclidean signature the action \eqref{MSM2action4} has an overall negative sign, which is surprisingly absent in the large-scale effective action of causal dynamical triangulations \cite{JA&DNC&JGS&JJ,JA&JGS&AG&JJ,JA&JGS&AG&JJ2,JA&AG&JJ&RL1,JA&AG&JJ&RL2,JA&AG&JJ&RL&JGS&TT,JA&JJ&RL4,JA&JJ&RL5,JA&JJ&RL6}. See \cite{JA&AG&JJ&RL3} for a plausible yet tentative explanation.} $G$ and $\Lambda$ are now the renormalized Newton and cosmological constants. The maximally symmetric extremum of the action \eqref{MSM2action4} for Euclidean signature is Euclidean de Sitter space, for which
\begin{equation}\label{dSvolprof}
V_{2}^{(\mathrm{EdS})}( t)=4\pi\ell_{\mathrm{dS}}^{2}\cos^{2}{\left(\frac{\omega t}{\ell_{\mathrm{dS}}}\right)}
\end{equation}
with $ t\in[-\pi\ell_{\mathrm{dS}}/2\omega,+\pi\ell_{\mathrm{dS}}/2\omega]$; the maximally symmetric extremum of the action \eqref{MSM2action4} for Lorentzian signature is Lorentzian de Sitter spacetime, for which
\begin{equation}\label{LdSvolprof}
V_{2}^{(\mathrm{LdS})}( t)=4\pi \ell_{\mathrm{dS}}^{2}\cosh^{2}{\left(\frac{\omega t}{\ell_{\mathrm{dS}}}\right)}
\end{equation}
with $ t\in(-\infty,+\infty)$. $\ell_{\mathrm{dS}}=\sqrt{1/\Lambda}$ is the de Sitter length. 

We first model the ensemble average number $\langle N_{2}^{\mathrm{SL}}(\tau)\rangle$ of spacelike $2$-simplices as a function of the discrete time coordinate $\tau$ on the basis of the spatial $2$-volumes $V_{2}^{(\mathrm{EdS})}( t)$ and $V_{2}^{(\mathrm{LdS)}}(t)$ given in equations \eqref{dSvolprof} and \eqref{LdSvolprof}. In particular, we derive a discrete analogue $\mathcal{N}_{2}^{\mathrm{SL}}(\tau)$ appropriate to causal triangulations of each of the spatial $2$-volumes $V_{2}^{(\mathrm{EdS})}( t)$ and $V_{2}^{(\mathrm{LdS)}}(t)$, and we subsequently perform a best fit of $\mathcal{N}_{2}^{\mathrm{SL}}(\tau)$ to $\langle N_{2}^{\mathrm{SL}}(\tau)\rangle$. 
Several authors have previously performed such a derivation in the case of Euclidean de Sitter space \cite{JA&AG&JJ&RL1,JA&AG&JJ&RL2,JA&AG&JJ&RL3,JA&AG&JJ&RL&JGS&TT,JA&JJ&RL3,JA&JJ&RL4,JA&JJ&RL5,JA&JJ&RL6,CA&SJC&JHC&PH&RKK&PZ,DB&JH2,JHC,JHC&JMM}; we adapt their techniques to the case of a portion of Lorentzian de Sitter spacetime. 
We begin by assuming a canonical finite-size scaling \emph{Ansatz} based on the double scaling limit 
\begin{equation}\label{FSSansatz}
V_{3}=\lim_{\substack{N_{3}\rightarrow\infty \\ a\rightarrow0}}C_{3}N_{3}a^{3}
\end{equation}
of the spacetime $3$-volume $V_{3}$: in the infinite-volume ($N_{3}\rightarrow\infty$) and continuum ($a\rightarrow0$) limits, the discrete spacetime $3$-volume $C_{3}N_{3}a^{3}$ approaches the constant value $V_{3}$. $C_{3}$ is the effective discrete spacetime $3$-volume of a single $3$-simplex. Evidence for the applicability of this \emph{Ansatz} to the scaling of $\langle N_{2}^{\mathrm{SL}}(\tau)\rangle$ is presented in \cite{JA&AG&JJ&RL3,JA&JJ&RL4,JA&JJ&RL5,JA&JJ&RL6,DB&JH2}. The motivation for this \emph{Ansatz} is as following: $V_{3}$ is the largest-scale physical observable present in our model, so, of all possible discrete observables, we expect the discrete spacetime 3-volume to scale canonically with $N_{3}$ and $a$. In appendix \ref{derivation1pt} we employ the finite-size scaling \emph{Ansatz} based on equation \eqref{FSSansatz} to derive the discrete analogue $\mathcal{N}_{2}^{\mathrm{SL}}(\tau)$ for each of the spatial $2$-volumes $V_{2}^{(\mathrm{EdS})}( t)$ and $V_{2}^{(\mathrm{LdS})}( t)$ restricted to the finite global time interval $[ t_{\mathrm{i}}, t_{\mathrm{f}}]$. In the case of Euclidean de Sitter space, we derive that
\begin{equation}\label{variantdiscretedSvolprofileappen}
\mathcal{N}_{2}^{\mathrm{SL}}(\tau)=\frac{\langle N_{3}^{(1,3)}\rangle}{\bar{s}_{0}\langle N_{3}^{(1,3)}\rangle^{1/3}}\frac{\cos^{2}{\left(\frac{\tau}{\bar{s}_{0}\langle N_{3}^{(3,1)}\rangle^{1/3}}\right)}}{\frac{\tau_{\mathrm{f}}-\tau_{\mathrm{i}}}{\bar{s}_{0}\langle N_{3}^{(1,3)}\rangle^{1/3}}+2\sin{\left(\frac{\tau_{\mathrm{f}}-\tau_{\mathrm{i}}}{\bar{s}_{0}\langle N_{3}^{(1,3)}\rangle^{1/3}}\right)}\cos{\left(\frac{\tau_{\mathrm{f}}+\tau_{\mathrm{i}}}{\bar{s}_{0}\langle N_{3}^{(1,3)}\rangle^{1/3}}\right)}},
\end{equation}
as previously determined in \cite{JHC&JMM}, and, in the case of Lorentzian de Sitter spacetime, we derive that
\begin{equation}\label{LdSdiscreteanalogue}
\mathcal{N}_{2}^{\mathrm{SL}}(\tau)=\frac{\langle N_{3}^{(1,3)}\rangle}{\bar{s}_{0}\langle N_{3}^{(1,3)}\rangle^{1/3}}\frac{\cosh^{2}{\left(\frac{\tau}{\bar{s}_{0}\langle N_{3}^{(1,3)}\rangle^{1/3}}\right)}}{\frac{\tau_{\mathrm{f}}-\tau_{\mathrm{i}}}{\bar{s}_{0}\langle N_{3}^{(1,3)}\rangle^{1/3}}+2\sinh{\left(\frac{\tau_{\mathrm{f}}-\tau_{\mathrm{i}}}{\bar{s}_{0}\langle N_{3}^{(1,3)}\rangle^{1/3}}\right)}\cosh{\left(\frac{\tau_{\mathrm{f}}+\tau_{\mathrm{i}}}{\bar{s}_{0}\langle N_{3}^{(1,3)}\rangle^{1/3}}\right)}}.
\end{equation}
$N_{3}^{(1,3)}$ is the number of $(1,3)$ $3$-simplices, 
\begin{equation}
\bar{s}_{0}=\frac{2^{1/3}(1+\xi)^{1/3}\ell_{\mathrm{dS}}}{\omega V_{3}^{1/3}}
\end{equation}
is a fit parameter, and $\xi$ is the ratio of $\langle N_{3}^{(2,2)}\rangle$ to $\langle N_{3}^{(1,3)}\rangle+\langle N_{3}^{(3,1)}\rangle$. We now perform best fits of $\mathcal{N}_{2}^{\mathrm{SL}}(\tau)$ to the measurements of $\langle N_{2}^{\mathrm{SL}}(\tau)\rangle$ 
following the procedure of \cite{JHC&JMM}. We report the value $\chi_{\mathrm{red}}^{2}$ of the $\chi^{2}$ per degree of freedom for each fit. 

\begin{figure}
\centering
\includegraphics[scale=1]{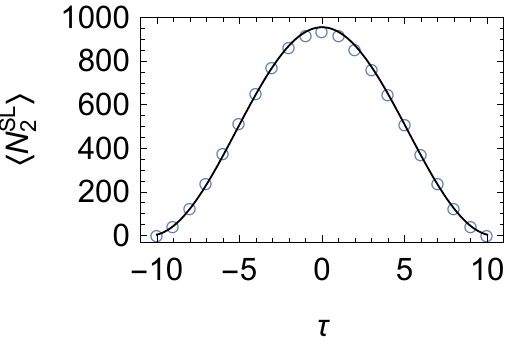}
\caption{Ensemble average number $\langle N_{2}^{\mathrm{SL}}\rangle$ of spacelike $2$-simplices as a function of the discrete time coordinate $\tau$ (blue circles) for $\bar{T}=21$, $\bar{N}_{3}=30850$, $k_{0}=1.00$, and $N_{2}^{\mathrm{SL}}(\mathsf{S}_{\mathrm{i}}^{2})=N_{2}^{\mathrm{SL}}(\mathsf{S}_{\mathrm{f}}^{2})=4$ (Euclidean-like ensemble $\mathcal{E}_{\mathrm{E}}$) overlain with the best fit discrete analogue $\mathcal{N}_{2}^{\mathrm{SL}}(\tau)$ (black line) of the spatial $2$-volume $V_{2}^{(\mathrm{EdS})}( t)$ as a function of the global time coordinate $t$ of Euclidean de Sitter space. $\chi^{2}_{\mathrm{red}}=79.91$.}
\label{volproffitT21V30K1IBFB4}
\end{figure}

To establish a point of comparison, we first consider the ensemble $\mathcal{E}_{\mathrm{E}}$ characterized by $\bar{T}=21$, $\bar{N}_{3}=30850$, $k_{0}=1.00$, and $N_{2}^{\mathrm{SL}}(\mathsf{S}_{\mathrm{i}}^{2})=N_{2}^{\mathrm{SL}}(\mathsf{S}_{\mathrm{f}}^{2})=4$, for which, as depicted in figure \ref{volproffitT21V30K1IBFB4}, $\langle N_{2}^{\mathrm{SL}}(\tau)\rangle$ exhibits the characteristic behavior of phase C. We display $\langle N_{2}^{\mathrm{SL}}(\tau)\rangle$ overlain with the best fit form of $\mathcal{N}_{2}^{\mathrm{SL}}(\tau)$, given in equation \eqref{variantdiscretedSvolprofileappen}, for the ensemble $\mathcal{E}_{\mathrm{E}}$ in figure \ref{volproffitT21V30K1IBFB4}. 
This fit of $\mathcal{N}_{2}^{\mathrm{SL}}(\tau)$ to $\langle N_{2}^{\mathrm{SL}}(\tau)\rangle$ is representative of the application of the above Euclidean model to measurements of $\langle N_{2}^{\mathrm{SL}}(\tau)\rangle$ \cite{JA&AG&JJ&RL1,JA&AG&JJ&RL2,JA&AG&JJ&RL3,JA&AG&JJ&RL&JGS&TT,JA&JJ&RL4,JA&JJ&RL5,JA&JJ&RL6,CA&SJC&JHC&PH&RKK&PZ,DB&JH2,JHC,JHC&JMM,RK}. Visually, $\mathcal{N}_{2}^{\mathrm{SL}}(\tau)$ fits $\langle N_{2}^{\mathrm{SL}}(\tau)\rangle$ quite satisfactorily. As measured by $\chi_{\mathrm{red}}^{2}$, the quality of the fit of $\mathcal{N}_{2}^{\mathrm{SL}}(\tau)$, given in equation \eqref{variantdiscretedSvolprofileappen}, to $\langle N_{2}^{\mathrm{SL}}(\tau)\rangle$ for this  Euclidean-like ensemble is comparable to the quality of previous such fits  \cite{JHC&JMM}. 

\begin{figure}[!ht]
\centering
\setlength{\unitlength}{\textwidth}
\begin{picture}(1,0.22)
\put(0,0.005){\includegraphics[scale=1]{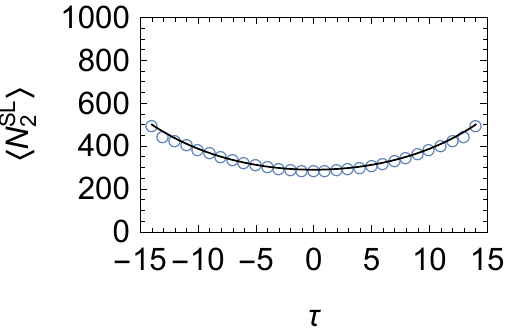}}
\put(0.34,0.005){\includegraphics[scale=1]{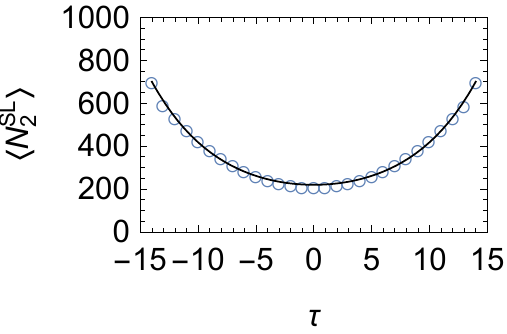}}
\put(0.67,0.005){\includegraphics[scale=1]{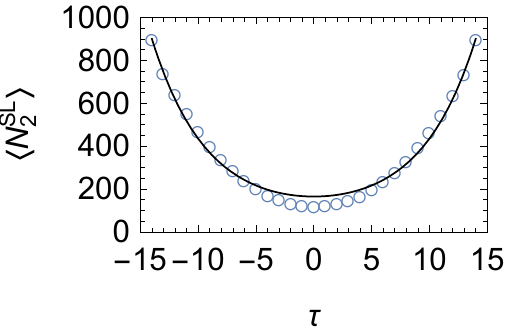}}
\put(0.18,-0.02){(a)}
\put(0.52,-0.02){(b)}
\put(0.85,-0.02){(c)}
\end{picture}
\caption{Ensemble average number $\langle N_{2}^{\mathrm{SL}}\rangle$ of spacelike $2$-simplices as a function of the discrete time coordinate $\tau$ (blue circles) for $\bar{T}=29$, $\bar{N}_{3}=30850$, and $k_{0}=1.00$ overlain with the best fit discrete analogue $\mathcal{N}_{2}^{\mathrm{SL}}(\tau)$ (black line) of the spatial $2$-volume $V_{2}^{(\mathrm{LdS})}( t)$ as a function of the global time coordinate $t$ of Lorentzian de Sitter spacetime. (a), $N_{2}^{\mathrm{SL}}(\mathsf{S}_{\mathrm{i}}^{2})=N_{2}^{\mathrm{SL}}(\mathsf{S}_{\mathrm{f}}^{2})=500$ $\chi^{2}_{\mathrm{red}}=169.86$. (b) $N_{2}^{\mathrm{SL}}(\mathsf{S}_{\mathrm{i}}^{2})=N_{2}^{\mathrm{SL}}(\mathsf{S}_{\mathrm{f}}^{2})=700$, $\chi^{2}_{\mathrm{red}}=143.44$. (c) $N_{2}^{\mathrm{SL}}(\mathsf{S}_{\mathrm{i}}^{2})=N_{2}^{\mathrm{SL}}(\mathsf{S}_{\mathrm{f}}^{2})=900$, $\chi^{2}_{\mathrm{red}}=1435.51$.}
\label{nonminnonminsame2fit}
\end{figure}

\begin{figure}[!ht]
\centering
\setlength{\unitlength}{\textwidth}
\begin{picture}(1,0.2)
\put(0.1,0.005){\includegraphics[scale=1]{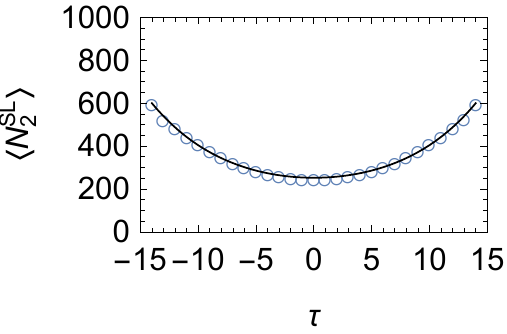}}
\put(0.5,0.005){\includegraphics[scale=1]{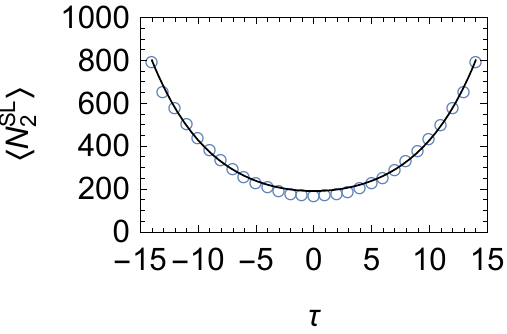}}
\put(0.28,-0.02){(a)}
\put(0.68,-0.02){(b)}
\end{picture}
\caption{Ensemble average number $\langle N_{2}^{\mathrm{SL}}\rangle$ of spacelike $2$-simplices as a function of the discrete time coordinate $\tau$ (blue circles) for $\bar{T}=29$, $\bar{N}_{3}=30850$, and $k_{0}=1.00$ overlain with the best fit discrete analogue $\mathcal{N}_{2}^{\mathrm{SL}}(\tau)$ (black line) of the spatial $2$-volume $V_{2}^{(\mathrm{LdS})}( t)$ as a function of the global time coordinate $t$ of Lorentzian de Sitter spacetime. (a) $N_{2}^{\mathrm{SL}}(\mathsf{S}_{\mathrm{i}}^{2})=N_{2}^{\mathrm{SL}}(\mathsf{S}_{\mathrm{f}}^{2})=600$, $\chi^{2}_{\mathrm{red}}=86.67$. (b) $N_{2}^{\mathrm{SL}}(\mathsf{S}_{\mathrm{i}}^{2})=N_{2}^{\mathrm{SL}}(\mathsf{S}_{\mathrm{f}}^{2})=800$, $\chi^{2}_{\mathrm{red}}=452.85$.}
\label{nonminnonmin2fit}
\end{figure}

We now test the hypothesis that a portion of Lorentzian de Sitter spacetime accurately describes the ensemble average number $\langle N_{2}^{\mathrm{SL}}(\tau)\rangle$ of spacelike $2$-simplices as a function of the discrete time coordinate $\tau$ for the Lorentzian-like ensembles represented in figures \ref{nonminnonminsame2} and \ref{nonminnonmin2}. 
We consider the five ensembles of causal triangulations represented in figures \ref{nonminnonminsame2} and \ref{nonminnonmin2} including the ensemble $\mathcal{E}_{\mathrm{L}}$.
For $\bar{T}=29$, $\bar{N}_{3}=30850$, $k_{0}=1.00$, we display $\langle N_{2}^{\mathrm{SL}}(\tau)\rangle$ overlain with the best fit form of $\mathcal{N}_{2}^{\mathrm{SL}}(\tau)$, given in equation \eqref{LdSdiscreteanalogue}, for $N_{2}^{\mathrm{SL}}(\mathsf{S}_{\mathrm{i}}^{2})=N_{2}^{\mathrm{SL}}(\mathsf{S}_{\mathrm{f}}^{2})=500$ in figure \ref{nonminnonminsame2fit}(a), for $N_{2}^{\mathrm{SL}}(\mathsf{S}_{\mathrm{i}}^{2})=N_{2}^{\mathrm{SL}}(\mathsf{S}_{\mathrm{f}}^{2})=700$ in figure \ref{nonminnonminsame2fit}(b), for $N_{2}^{\mathrm{SL}}(\mathsf{S}_{\mathrm{i}}^{2})=N_{2}^{\mathrm{SL}}(\mathsf{S}_{\mathrm{f}}^{2})=900$ in figure \ref{nonminnonminsame2fit}(c),
for $N_{2}^{\mathrm{SL}}(\mathsf{S}_{\mathrm{i}}^{2})=N_{2}^{\mathrm{SL}}(\mathsf{S}_{\mathrm{f}}^{2})=600$ in figure \ref{nonminnonmin2fit}(a), and for $N_{2}^{\mathrm{SL}}(\mathsf{S}_{\mathrm{i}}^{2})=N_{2}^{\mathrm{SL}}(\mathsf{S}_{\mathrm{f}}^{2})=800$ in figure \ref{nonminnonmin2fit}(b).
Visually, $\mathcal{N}_{2}^{\mathrm{SL}}(\tau)$ again fits $\langle N_{2}^{\mathrm{SL}}(\tau)\rangle$ quite satisfactorily. As measured by $\chi_{\mathrm{red}}^{2}$, the quality of the fits of $\mathcal{N}_{2}^{\mathrm{SL}}(\tau)$, given in equation \eqref{LdSdiscreteanalogue}, to $\langle N_{2}^{\mathrm{SL}}(\tau)\rangle$ for these five Lorentzian-like ensembles is comparable to the quality of the fit of $\mathcal{N}_{2}^{\mathrm{SL}}(\tau)$, given in equation \eqref{variantdiscretedSvolprofileappen}, to $\langle N_{2}^{\mathrm{SL}}(\tau)\rangle$ for Euclidean-like ensembles \cite{JHC&JMM}. In particular, these fits for the ensembles $\mathcal{E}_{\mathrm{E}}$ and $\mathcal{E}_{\mathrm{L}}$ have nearly equivalent $\chi_{\mathrm{red}}^{2}$ values, motivating our choice to compare the ensembles $\mathcal{E}_{\mathrm{E}}$ and $\mathcal{E}_{\mathrm{L}}$. There is a systematic trend in the $\chi_{\mathrm{red}}^{2}$ values for these five Lorentzian-like ensembles: $\chi_{\mathrm{red}}^{2}$ is minimal for $N_{2}^{\mathrm{SL}}(\mathsf{S}_{\mathrm{i}}^{2})=N_{2}^{\mathrm{SL}}(\mathsf{S}_{\mathrm{f}}^{2})=600$ and increases monotonically for both smaller and larger values of $N_{2}^{\mathrm{SL}}(\mathsf{S}_{\mathrm{i}}^{2})=N_{2}^{\mathrm{SL}}(\mathsf{S}_{\mathrm{f}}^{2})$. Cooperman and Miller found the same type of trend for ensembles with different numbers $\bar{T}$ of time slices at fixed number $\bar{N}_{3}$ of $3$-simplices, coupling $k_{0}$, and numbers $N_{2}^{\mathrm{SL}}(\mathsf{S}_{\mathrm{i}}^{2})$ and $N_{2}^{\mathrm{SL}}(\mathsf{S}_{\mathrm{f}}^{2})$ of initial and final spacelike $2$-simplices \cite{JHC&JMM}. These trends likely stem from either undiagnosed finite-size effects or incomplete modeling. We touch on finite-size scaling analyses of transition amplitudes at the end of this section, and Cooperman and Houthoff perform a first investigation of systematic modeling issues in a forthcoming paper \cite{JHC&WH}.

We now extend our model to include linear gravitational perturbations $v_{2}(t)$ propagating on either Euclidean de Sitter space or Lorentzian de Sitter spacetime. In the path integral formalism one computes the connected $2$-point function $\mathbb{E}_{\mathrm{EdS}}[v_{2}(t)\,v_{2}(t')]$ of perturbations $v_{2}(t)$ about Euclidean de Sitter space as
\begin{equation}\label{EdS2ptdef}
\mathbb{E}_{\mathrm{EdS}}[v_{2}(t)\,v_{2}(t')]=\frac{\int\mathrm{d}\mu(v_{2})\,v_{2}( t)\,v_{2}( t')\,e^{-S_{\mathrm{cl}}[v_{2}]/\hbar}}{\int\mathrm{d}\mu(v_{2})\,e^{-S_{\mathrm{cl}}[v_{2}]/\hbar}},
\end{equation}
in which $S_{\mathrm{cl}}[v_{2}]$ is the action \eqref{MSM2action4} in Euclidean signature for the spatial $2$-volume $V_{2}(t)$ perturbed by $v_{2}(t)$ about $V_{2}^{(\mathrm{EdS})}(t)$, and the connected $2$-point function $\mathbb{E}_{\mathrm{LdS}}[v_{2}(t)\,v_{2}(t')]$ of perturbations $v_{2}(t)$ about Lorentzian de Sitter spacetime as
\begin{equation}\label{LdS2ptdef}
\mathbb{E}_{\mathrm{LdS}}[v_{2}(t)\,v_{2}(t')]=\frac{\int\mathrm{d}\mu(v_{2})\,v_{2}( t)\,v_{2}( t')\,e^{iS_{\mathrm{cl}}[v_{2}]/\hbar}}{\int\mathrm{d}\mu(v_{2})\,e^{iS_{\mathrm{cl}}[v_{2}]/\hbar}},
\end{equation}
in which $S_{\mathrm{cl}}[v_{2}]$ is the action \eqref{MSM2action4} in Lorentzian signature for the spatial $2$-volume $V_{2}(t)$ perturbed by $v_{2}(t)$ about $V_{2}^{(\mathrm{LdS})}(t)$. Expanding the action \eqref{MSM2action4} in Euclidean signature to second order in $v_{2}(t)$, assuming that $V_{2}^{\mathrm{(EdS)}}(t)\gg v_{2}(t)$, we find that
\begin{eqnarray}\label{modelaction2ndorder}
S_{\mathrm{cl}}[v_{2}]&=&S_{\mathrm{cl}}[V_{2}^{(\mathrm{EdS})}]-\frac{1}{64\pi^{2}G\ell_{\mathrm{dS}}^{3}}\int_{\tilde{t}_{\mathrm{i}}}^{\tilde{t}_{\mathrm{f}}}\mathrm{d}\tilde{t}\,v_{2}(\tilde{t})\sec^{2}{\tilde{t}}\left[\frac{\mathrm{d}^{2}}{\mathrm{d}\tilde{t}^{2}}+2\tan{\tilde{t}}\frac{\mathrm{d}}{\mathrm{d}\tilde{t}}+2\sec^{2}{\tilde{t}}\right]v_{2}(\tilde{t})\nonumber\\ &&\qquad+O\left[\left(v_{2}\right)^{3}\right],
\end{eqnarray}
for $\tilde{ t}=\omega t/\ell_{\mathrm{dS}}$. The terms of first order in $v_{2}(t)$ vanish because $V_{2}^{(\mathrm{EdS})}(t)$ is an extremum of the action \eqref{MSM2action4} in Euclidean signature. Expanding the action \eqref{MSM2action4} in Lorentzian signature to second order in $v_{2}(t)$, assuming that $V_{2}^{\mathrm{(LdS)}}(t)\gg v_{2}(t)$, we find that
\begin{eqnarray}
S_{\mathrm{cl}}[v_{2}]&=&S_{\mathrm{cl}}[V_{2}^{(\mathrm{LdS})}]+\frac{1}{64\pi^{2}G\ell_{\mathrm{dS}}^{3}}\int_{\tilde{t}_{\mathrm{i}}}^{\tilde{t}_{\mathrm{f}}}\mathrm{d}\tilde{ t}\,v_{2}(\tilde{t})\sech^{2}{\tilde{ t}}\left[\frac{\mathrm{d}^{2}}{\mathrm{d}\tilde{ t}^{2}}-2\tanh{\tilde{ t}}\frac{\mathrm{d}}{\mathrm{d}\tilde{ t}}-2\sech^{2}{\tilde{ t}}\right]v_{2}(\tilde{t})\nonumber\\ &&\qquad+O\left[\left(v_{2}\right)^{3}\right]
\end{eqnarray}
for $\tilde{ t}=\omega t/\ell_{\mathrm{dS}}$. The terms of first order in $v_{2}(t)$ vanish because $V_{2}^{(\mathrm{LdS})}(t)$ is an extremum of the action \eqref{MSM2action4} in Lorentzian signature. A standard calculation now gives that
\begin{equation}
\mathbb{E}[v_{2}( t)\,v_{2}( t')]=\left[\frac{1}{\hbar}\mathscr{M}( t, t')\right]^{-1},
\end{equation}
in which
\begin{equation}
\mathscr{M}( t, t')=\frac{\delta^{2}S_{\mathrm{cl}}[v_{2}]}{\delta v_{2}( t)\,\delta v_{2}( t')}\bigg|_{\substack{v_{2}( t)=0 \\ v_{2}( t')=0}}
\end{equation}
is the van Vleck-Morette determinant. For perturbations $v_{2}(t)$ about the spatial $2$-volume $V_{2}^{(\mathrm{EdS})}(t)$ of Euclidean de Sitter space,
\begin{equation}\label{vVMdEdS}
\mathscr{M}( t, t')=\frac{1}{64\pi^{2}G\ell_{\mathrm{dS}}^{3}}\sec^{2}{\tilde{t}}\left[\frac{\mathrm{d}^{2}}{\mathrm{d}\tilde{t}^{2}}+2\tan{\tilde{t}}\frac{\mathrm{d}}{\mathrm{d}\tilde{t}}+2\sec^{2}{\tilde{t}}\right],
\end{equation}
and, for perturbations $v_{2}(t)$ about the spatial $2$-volume $V_{2}^{(\mathrm{LdS})}(t)$ of Lorentzian de Sitter spacetime,
\begin{equation}\label{vVMdLdS}
\mathscr{M}( t, t')=\frac{1}{64\pi^{2}G\ell_{\mathrm{dS}}^{3}}\sech^{2}{\tilde{ t}}\left[\frac{\mathrm{d}^{2}}{\mathrm{d}\tilde{ t}^{2}}-2\tanh{\tilde{ t}}\frac{\mathrm{d}}{\mathrm{d}\tilde{ t}}-2\sech^{2}{\tilde{ t}}\right].
\end{equation}
One can show moreover that
\begin{equation}\label{vVMdetmodesum}
\mathscr{M}( t, t')=\sum_{j=1}^{\infty}\mu_{j}\,\nu_{j}( t)\,\nu_{j}( t')
\end{equation}
in which $\nu_{j}( t)$ are the eigenfunctions of the operator $\mathscr{M}( t, t')$ with associated eigenvalues $\mu_{j}$ satisfying the integral constraint
\begin{equation}\label{eigenfunctionconstraint}
\int_{ t_{\mathrm{i}}}^{ t_{\mathrm{f}}}\mathrm{d} t\,\omega\nu_{j}( t)=0
\end{equation}
and the boundary conditions $\nu_{j}( t_{\mathrm{i}})=0$ and $\nu_{j}( t_{\mathrm{f}})=0$. Accordingly, 
\begin{equation}\label{2ptmodesum}
\mathbb{E}[v_{2}( t)\,v_{2}( t')]=\sum_{j=1}^{\infty}\frac{\hbar}{\mu_{j}}\,\nu_{j}( t)\,\nu_{j}( t')
\end{equation}
assuming that $\mu_{j}\neq0$ for all $j$, which holds in the cases under consideration. 


We next model the ensemble average connected $2$-point function $\langle n_{2}^{\mathrm{SL}}(\tau)\,n_{2}^{\mathrm{SL}}(\tau')\rangle$ of deviations $n_{2}^{\mathrm{SL}}(\tau)$ in the number $N_{2}^{\mathrm{SL}}(\tau)$ of spacelike $2$-simplices as a function of the discrete time coordinate $\tau$ on the basis of the $2$-point functions $\mathbb{E}_{\mathrm{EdS}}[v_{2}(t)\,v_{2}(t')]$ and $\mathbb{E}_{\mathrm{LdS}}[v_{2}(t)\,v_{2}(t')]$ given in equations \eqref{EdS2ptdef} and \eqref{LdS2ptdef}. 
In particular, we derive a discrete analogue $\mathsf{n}_{2}^{\mathrm{SL}}(\tau)\,\mathsf{n}_{2}^{\mathrm{SL}}(\tau')$ appropriate to causal triangulations of each of the $2$-point functions 
$\mathbb{E}_{\mathrm{EdS}}[v_{2}(t)\,v_{2}(t')]$ and $\mathbb{E}_{\mathrm{LdS}}[v_{2}(t)\,v_{2}(t')]$, and we subsequently perform a fit of $\mathsf{n}_{2}^{\mathrm{SL}}(\tau)\,\mathsf{n}_{2}^{\mathrm{SL}}(\tau')$ to $\langle n_{2}^{\mathrm{SL}}(\tau)\,n_{2}^{\mathrm{SL}}(\tau')\rangle$. Ambj\o rn \emph{et al} \cite{JA&AG&JJ&RL1,JA&AG&JJ&RL2} and Cooperman \cite{JHC} have previously performed such a derivation in the case of Euclidean de Sitter space in $3+1$ dimensions; we adapt their techniques to the cases of Euclidean de Sitter space in $2+1$ dimensions and a portion of Lorentzian de Sitter spacetime in $2+1$ dimensions. We again assume the finite-size scaling \emph{Ansatz} based on equation \eqref{FSSansatz}.  
In appendix \ref{derivation2pt} we employ the \emph{Ansatz} based on equation \eqref{FSSansatz} to derive the discrete analogue $\mathsf{n}_{2}^{\mathrm{SL}}(\tau)\,\mathsf{n}_{2}^{\mathrm{SL}}(\tau')$ for each of the $2$-point functions $\mathbb{E}_{\mathrm{EdS}}[v_{2}( t)\, v_{2}( t')]$ and $\mathbb{E}_{\mathrm{LdS}}[v_{2}( t)\, v_{2}( t')]$. Specifically, we derive $\mathsf{n}_{2}^{\mathrm{SL}}(\tau)\,\mathsf{n}_{2}^{\mathrm{SL}}(\tau')$ in the form of equation \eqref{2ptmodesum}, determining the eigenvectors $\nu_{j}(\tau)$ and associated eigenvalues $\mu_{j}$ of $\mathsf{n}_{2}^{\mathrm{SL}}(\tau)\,\mathsf{n}_{2}^{\mathrm{SL}}(\tau')$. We now perform fits of the eigenvectors $\nu_{j}(\tau)$ to the eigenvectors $\eta_{j}(\tau)$ and of the eigenvalues $\mu_{j}$ to the eigenvalues $\lambda_{j}$. Once the best fit of $\mathcal{N}_{2}^{\mathrm{SL}}(\tau)$ to $\langle N_{2}^{\mathrm{SL}}(\tau)\rangle$ fixes the fit parameter $\bar{s}_{0}$, there is in fact no fitting to perform aside from a single overall rescaling of the eigenvalues corresponding to the value of $1/64\pi^{2}\hbar G\ell_{\mathrm{dS}}^{3}$. Employing this value of $\bar{s}_{0}$ accords with our treatment of $v_{2}(t)$ as a perturbation. 

\begin{figure}
\centering
\includegraphics[width=\linewidth]{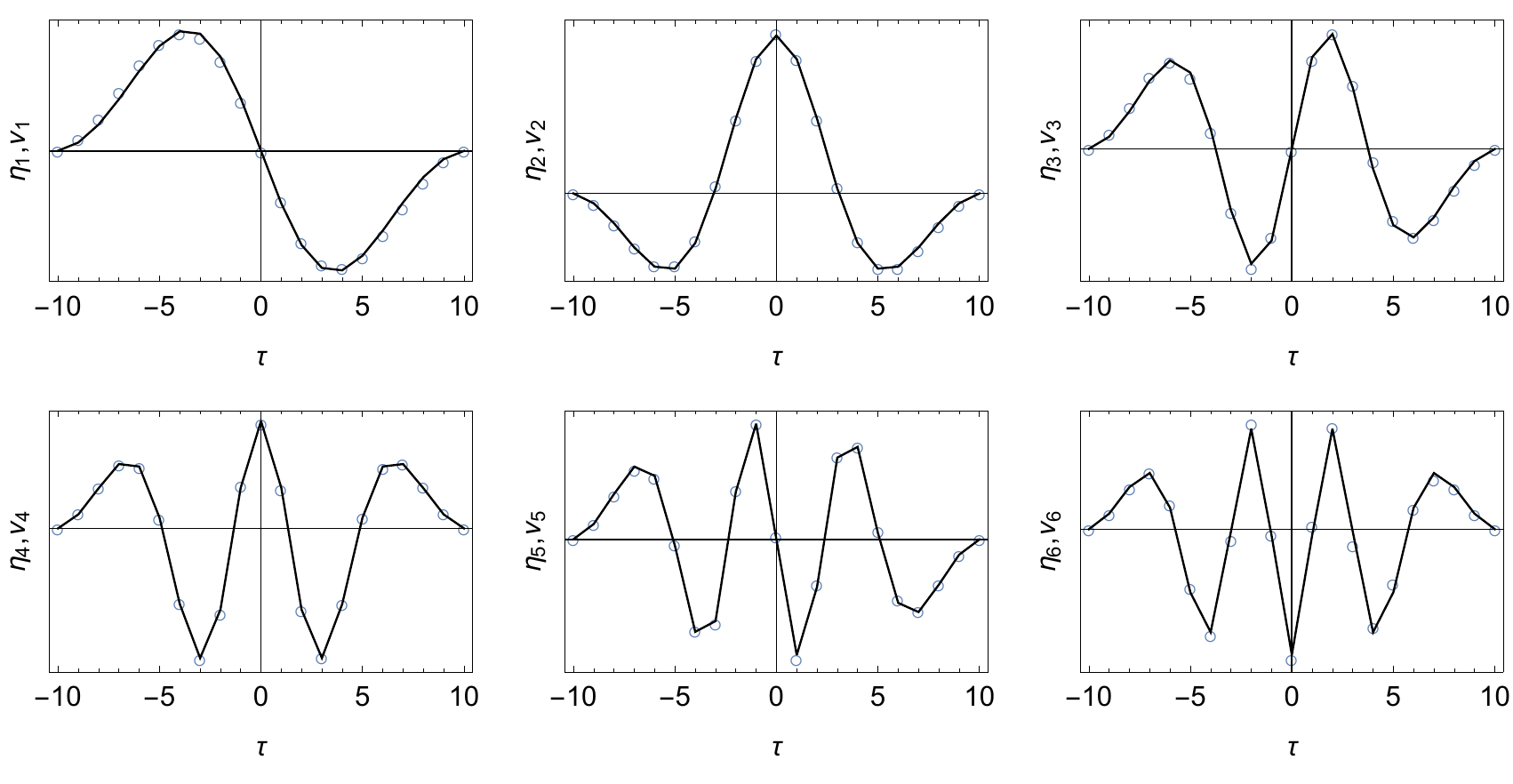}
\caption{First six eigenvectors $\eta_{j}(\tau)$ (blue circles) of the ensemble average connected $2$-point function $\langle n_{2}^{\mathrm{SL}}(\tau)\,n_{2}^{\mathrm{SL}}(\tau')\rangle$ of deviations $n_{2}^{\mathrm{SL}}(\tau)$ in the number of spacelike $2$-simplices as a function of the discrete time coordinate $\tau$ for $\bar{T}=21$, $\bar{N}_{3}=30850$, $k_{0}=1.00$, and $N_{2}^{\mathrm{SL}}(\mathsf{S}_{\mathrm{i}}^{2})=N_{2}^{\mathrm{SL}}(\mathsf{S}_{\mathrm{f}}^{2})=4$ (Euclidean-like ensemble $\mathcal{E}_{\mathrm{E}}$) overlain with the eigenvectors $\nu_{j}(\tau)$ (black lines) of the discrete analogue $\mathsf{n}_{2}^{\mathrm{SL}}(\tau)\,\mathsf{n}_{2}^{\mathrm{SL}}(\tau')$ of the connected $2$-point function $\mathbb{E}_{\mathrm{EdS}}[v_{2}( t)\,v_{2}( t')]$ of perturbations $v_{2}(t)$ in the spatial $2$-volume $V_{2}^{(\mathrm{EdS})}( t)$ as a function of the global time coordinate $ t$ of Euclidean de Sitter space.}
\label{eigenvectorfitE}
\end{figure}

\begin{figure}
\centering
\includegraphics[scale=1]{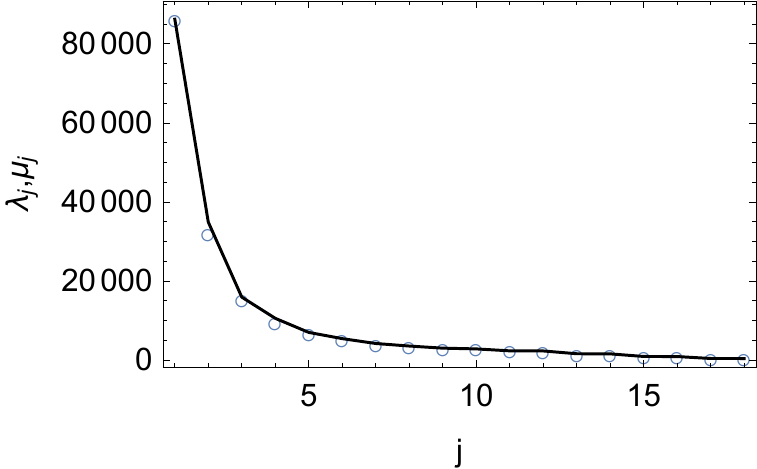}
\caption{Eigenvalues $\lambda_{j}$ (blue circles) of the ensemble average connected $2$-point function $\langle n_{2}^{\mathrm{SL}}(\tau)\,n_{2}^{\mathrm{SL}}(\tau')\rangle$ of deviations $n_{2}^{\mathrm{SL}}(\tau)$ in the number of spacelike $2$-simplices as a function of the discrete time coordinate $\tau$ for $\bar{T}=21$, $\bar{N}_{3}=30850$, $k_{0}=1.00$, and $N_{2}^{\mathrm{SL}}(\mathsf{S}_{\mathrm{i}}^{2})=N_{2}^{\mathrm{SL}}(\mathsf{S}_{\mathrm{f}}^{2})=4$ (Euclidean-like ensemble $\mathcal{E}_{\mathrm{E}}$) overlain with the eigenvalues $\mu_{j}$ (black lines) of the discrete analogue $\mathsf{n}_{2}^{\mathrm{SL}}(\tau)\,\mathsf{n}_{2}^{\mathrm{SL}}(\tau')$ of the connected $2$-point function $\mathbb{E}_{\mathrm{EdS}}[v_{2}( t)\,v_{2}( t')]$ of perturbations $v_{2}(t)$ in the spatial $2$-volume $V_{2}^{(\mathrm{EdS})}( t)$ as a function of the global time coordinate $ t$ of Euclidean de Sitter space.}
\label{eigenvaluefitE}
\end{figure}

To establish a point of comparison, we first consider the ensemble $\mathcal{E}_{\mathrm{E}}$ characterized by $\bar{T}=21$, $\bar{N}_{3}=30850$, $k_{0}=1.00$, and $N_{2}^{\mathrm{SL}}(\mathsf{S}_{\mathrm{i}}^{2})=N_{2}^{\mathrm{SL}}(\mathsf{S}_{\mathrm{f}}^{2})=4$, for which, as depicted in figure \ref{eigenvectorfitE}, $\langle n_{2}^{\mathrm{SL}}(\tau)\,n_{2}^{\mathrm{SL}}(\tau')\rangle$ exhibits the characteristic behavior of phase C. We display the first six eigenvectors $\eta_{j}(\tau)$ of $\langle n_{2}^{\mathrm{SL}}(\tau)\,n_{2}^{\mathrm{SL}}(\tau')\rangle$ overlain with the corresponding eigenvectors $\nu_{j}(\tau)$ of the discrete analogue $\mathsf{n}_{2}^{\mathrm{SL}}(\tau)\,\mathsf{n}_{2}^{\mathrm{SL}}(\tau)$ of $\mathbb{E}_{\mathrm{EdS}}[v_{2}(t)\,v_{2}(t')]$ in figure \ref{eigenvectorfitE}. 
We display the eigenvalues $\lambda_{j}$ of $\langle n_{2}^{\mathrm{SL}}(\tau)\,n_{2}^{\mathrm{SL}}(\tau')\rangle$ overlain with the corresponding eigenvalues $\mu_{j}$ of $\mathsf{n}_{2}^{\mathrm{SL}}(\tau)\,\mathsf{n}_{2}^{\mathrm{SL}}(\tau)$ in figure \ref{eigenvaluefitE}. 
The fits of $\nu_{j}(\tau)$ to $\eta_{j}(t)$ and of $\mu_{j}$ to $\lambda_{j}$ are representative of the application of the above Euclidean model to measurements of $\langle n_{2}^{\mathrm{SL}}(\tau)\,n_{2}^{\mathrm{SL}}(\tau')\rangle$ \cite{JA&AG&JJ&RL1,JA&AG&JJ&RL2,JHC}. Clearly, this model provides an accurate description of the connected $2$-point function $\langle n_{2}^{\mathrm{SL}}(\tau)\,n_{2}^{\mathrm{SL}}(\tau')\rangle$ for the ensemble $\mathcal{E}_{\mathrm{E}}$. 

\begin{figure}
\centering
\includegraphics[width=\linewidth]{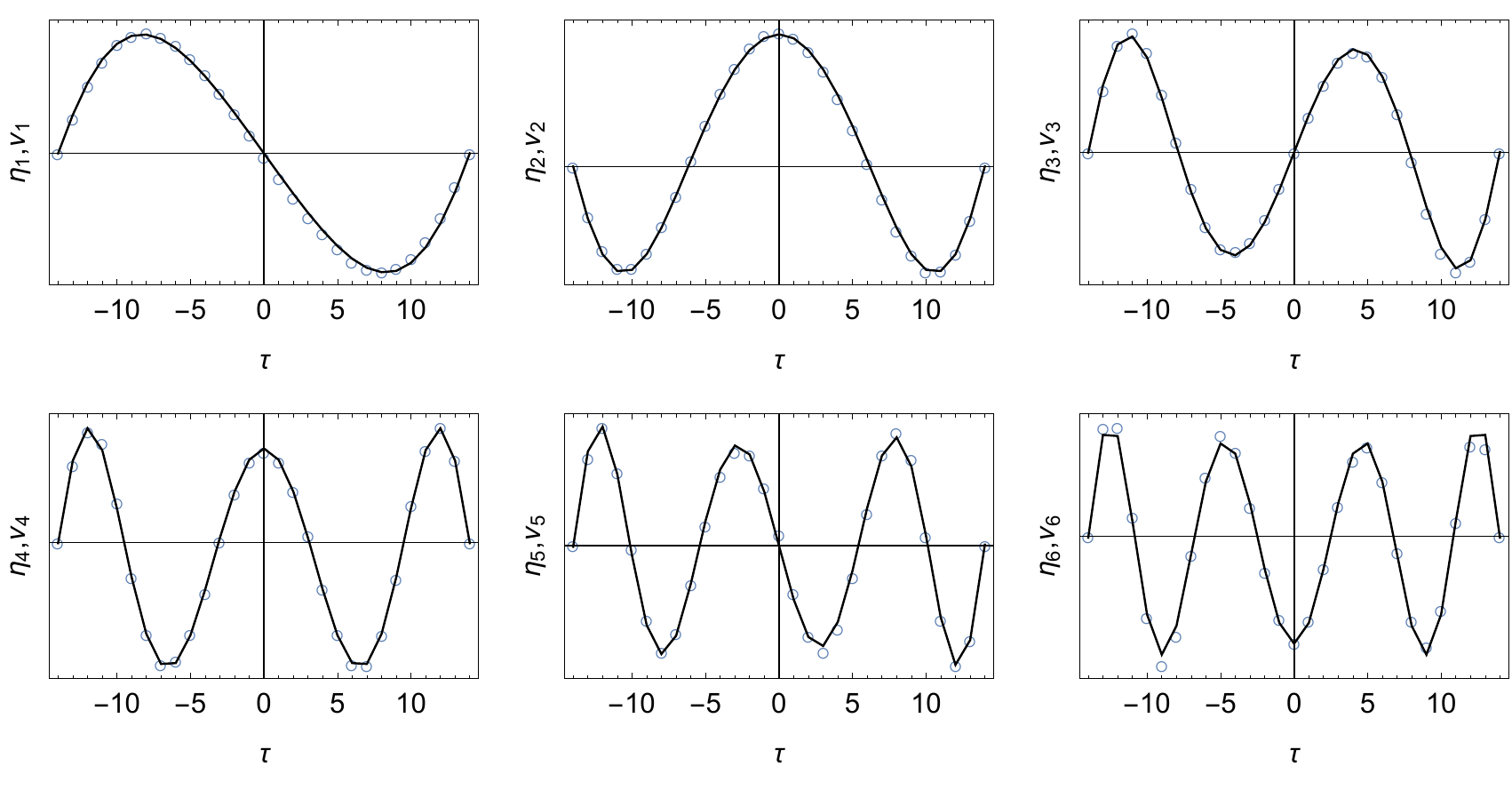}
\caption{First six eigenvectors $\eta_{j}(\tau)$ (blue circles) of the ensemble average connected $2$-point function $\langle n_{2}^{\mathrm{SL}}(\tau)\,n_{2}^{\mathrm{SL}}(\tau')\rangle$ of deviations $n_{2}^{\mathrm{SL}}(\tau)$ in the number of spacelike $2$-simplices as a function of the discrete time coordinate $\tau$ for $\bar{T}=29$, $\bar{N}_{3}=30850$, $k_{0}=1.00$, and $N_{2}^{\mathrm{SL}}(\mathsf{S}_{\mathrm{i}}^{2})=N_{2}^{\mathrm{SL}}(\mathsf{S}_{\mathrm{f}}^{2})=600$ (Lorentzian-like ensemble $\mathcal{E}_{\mathrm{L}}$) overlain with the eigenvectors $\nu_{j}(\tau)$ (black lines) of the discrete analogue $\mathsf{n}_{2}^{\mathrm{SL}}(\tau)\,\mathsf{n}_{2}^{\mathrm{SL}}(\tau')$ of the connected $2$-point function $\mathbb{E}_{\mathrm{LdS}}[v_{2}( t)\,v_{2}( t')]$ of perturbations $v_{2}(t)$ in the spatial $2$-volume $V_{2}^{(\mathrm{LdS})}( t)$ as a function of the global time coordinate $ t$ of Lorentzian de Sitter space.}
\label{eigenvectorfitL}
\end{figure}

\begin{figure}
\centering
\includegraphics[scale=1]{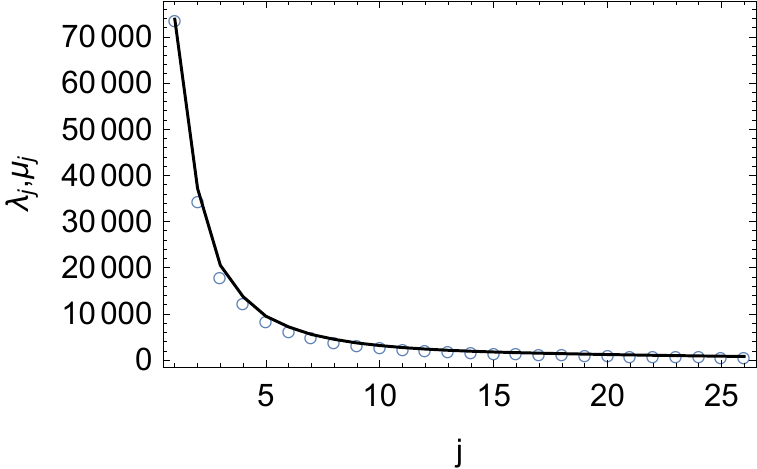}
\caption{Eigenvalues $\lambda_{j}$ (blue circles) of the ensemble average connected $2$-point function $\langle n_{2}^{\mathrm{SL}}(\tau)\,n_{2}^{\mathrm{SL}}(\tau')\rangle$ of deviations $n_{2}^{\mathrm{SL}}(\tau)$ in the number of spacelike $2$-simplices as a function of the discrete time coordinate $\tau$ for $\bar{T}=29$, $\bar{N}_{3}=30850$, $k_{0}=1.00$, and $N_{2}^{\mathrm{SL}}(\mathsf{S}_{\mathrm{i}}^{2})=N_{2}^{\mathrm{SL}}(\mathsf{S}_{\mathrm{f}}^{2})=600$ (Lorentzian-like ensemble $\mathcal{E}_{\mathrm{L}}$) overlain with the eigenvalues $\mu_{j}$ (black lines) of the discrete analogue $\mathsf{n}_{2}^{\mathrm{SL}}(\tau)\,\mathsf{n}_{2}^{\mathrm{SL}}(\tau')$ of the connected $2$-point function $\mathbb{E}_{\mathrm{LdS}}[v_{2}( t)\,v_{2}( t')]$ of perturbations $v_{2}(t)$ in the spatial $2$-volume $V_{2}^{(\mathrm{LdS})}( t)$ as a function of the global time coordinate $ t$ of Lorentzian de Sitter space.}
\label{eigenvaluefitL}
\end{figure}

We now test the hypothesis that the connected $2$-point function $\mathbb{E}_{\mathrm{LdS}}[v_{2}( t)\,v_{2}( t')]$ of linear gravitational perturbations $v_{2}( t)$ propagating on Lorentzian de Sitter spacetime accurately describes the shape of the ensemble average connected $2$-point function $\langle n_{2}^{\mathrm{SL}}(\tau)\,n_{2}^{\mathrm{SL}}(\tau')\rangle$ for the Lorentzian-like ensembles represented in figures \ref{nonminnonminsame2} and \ref{nonminnonmin2}. 
We consider only the ensemble $\mathcal{E}_{\mathrm{L}}$ characterized by $\bar{T}=29$, $\bar{N}_{3}=30850$, $k_{0}=1.00$, and $N_{2}^{\mathrm{SL}}(\mathsf{S}_{\mathrm{i}}^{2})=N_{2}^{\mathrm{SL}}(\mathsf{S}_{\mathrm{f}}^{2})=600$. We display the first six eigenvectors $\eta_{j}(\tau)$ of $\langle n_{2}^{\mathrm{SL}}(\tau)\,n_{2}^{\mathrm{SL}}(\tau')\rangle$ overlain with the corresponding eigenvectors $\nu_{j}(\tau)$ of the discrete analogue $\mathsf{n}_{2}^{\mathrm{SL}}(\tau)\,\mathsf{n}_{2}^{\mathrm{SL}}(\tau)$ of $\mathbb{E}_{\mathrm{LdS}}[v_{2}(t)\,v_{2}(t')]$ in figure \ref{eigenvectorfitL}. 
We display the eigenvalues $\lambda_{j}$ of $\langle n_{2}^{\mathrm{SL}}(\tau)\,n_{2}^{\mathrm{SL}}(\tau')\rangle$ overlain with the corresponding eigenvalues $\mu_{j}$ of $\mathsf{n}_{2}^{\mathrm{SL}}(\tau)\,\mathsf{n}_{2}^{\mathrm{SL}}(\tau)$ in figure \ref{eigenvaluefitL}.
Clearly, the above Lorentzian model provides an accurate description of the connected $2$-point function $\langle n_{2}^{\mathrm{SL}}(\tau)\,n_{2}^{\mathrm{SL}}(\tau')\rangle$ for the ensemble $\mathcal{E}_{\mathrm{L}}$.

These analyses, straightforwardly interpreted, provide evidence supporting the conjecture of Cooperman and Miller: a portion of Lorentzian de Sitter spacetime accurately describes the shape of $\langle N_{2}^{\mathrm{SL}}(\tau)\rangle$, and the connected $2$-point function of linear perturbations propagating on Lorentzian de Sitter spacetime accurately describes the shape of $\langle n_{2}^{\mathrm{SL}}(\tau)\,n_{2}^{\mathrm{SL}}(\tau')\rangle$ for the Lorentzian-like ensembles represented in figures \ref{nonminnonminsame2} and \ref{nonminnonmin2}. Nevertheless, we proffer an even more straightforward explanation of these results in section \ref{argumentrefutation}, casting serious doubt on the conjecture of Cooperman and Miller. 

Ideally, we would extend our analysis of the above modeling of $\langle N_{2}^{\mathrm{SL}}(\tau)\rangle$ and $\langle n_{2}^{\mathrm{SL}}(\tau)\,n_{2}^{\mathrm{SL}}(\tau')\rangle$ in two directions. First, we would perform a finite-size scaling analysis in which we consider ensembles of causal triangulations characterized by increasing numbers $\bar{N}_{3}$ of $3$-simplices---and commensurately increasing numbers $\bar{T}$ of time slices and $N_{2}^{\mathrm{SL}}(\mathsf{S}_{\mathrm{i}}^{2})$ and $N_{2}^{\mathrm{SL}}(\mathsf{S}_{\mathrm{f}}^{2})$ of initial and final boundary spacelike $2$-simplices---to extrapolate the accuracy of our modeling towards the infinite-volume limit. Such a finite-size scaling analysis is more difficult to perform in the context of transition amplitudes: 
the manner in which one must commensurately increase $N_{2}^{\mathrm{SL}}(\mathsf{S}_{\mathrm{i}}^{2})$, $N_{2}^{\mathrm{SL}}(\mathsf{S}_{\mathrm{f}}^{2})$, and $\bar{T}$ with $\bar{N}_{3}$ to consider transition amplitudes related by the finite-size scaling \emph{Ansatz} based on equation \eqref{FSSansatz} is nontrivial. 
For this reason we have not yet performed any scaling analyses of the transition amplitudes; rather, we rely on the similarities of our numerical measurements to those of previous studies as justification for our use of the finite-size scaling \emph{Ansatz} based on equation \eqref{FSSansatz}. 
Second, we would consider models based on departures from  Einstein gravity---for instance, Ho\v{r}ava-Lifshitz or higher-order gravity---to assess our model's accuracy. Cooperman and Houthoff perform such an analysis, though only for Euclidean-like ensembles, in a forthcoming paper \cite{JHC&WH}.

\section{Argument and refutation}\label{argumentrefutation}

Extraordinary claims require extraordinary evidence. The conjecture of Cooperman and Miller constitutes an extraordinary claim, but we now argue that the analyses presented in section \ref{analysissupport} of the measurements presented in section \ref{evidenceconjecture} do not furnish extraordinary evidence. We offer an alternative explanation of these measurements and their analysis, one much more plausible as well as much more mundane. 



We based the analyses of section \ref{analysissupport} on a minisuperspace truncation of $(2+1)$-dimensional Einstein gravity with either Euclidean de Sitter space or Lorentzian de Sitter spacetime as its ground state. As we presented this model in section \ref{analysissupport}, we did not incorporate with sufficient care the setting 
of our numerical simulations of causal triangulations. Recall from section \ref{CDT} that we run a given simulation at fixed number $\bar{N}_{3}$ of $3$-simplices and at fixed numbers $N_{2}^{\mathrm{SL}}(\mathsf{S}_{\mathrm{i}}^{2})$ and $N_{2}^{\mathrm{SL}}(\mathsf{S}_{\mathrm{f}}^{2})$ of initial and final boundary spacelike $2$-simplices.\footnote{We also fix the number $\bar{T}$ of time slices; however, our model allows for an arbitrary lapse---the constant $\omega$, which propagates into the fit parameter $\bar{s}_{0}$---so we do not impose a constraint associated with fixed $\bar{T}$.} We accounted for these constraints by normalizing $V_{2}^{(\mathrm{dS})}(t)$ to $V_{3}$ and $\mathcal{N}_{2}^{\mathrm{SL}}(\tau)$ to $\bar{N}_{3}$ in the derivation of appendix \ref{derivation1pt} and by enforcing boundary conditions on $\mathcal{N}_{2}^{\mathrm{SL}}(\tau)$ in the best fit to $\langle N_{2}^{\mathrm{SL}}(\tau)\rangle$. We did not, however, explicitly include constraints implementing a fixed spacetime $3$-volume $V_{3}$ and fixed initial and final spatial $2$-volumes $V_{2}(t_{\mathrm{i}})$ and $V_{2}(t_{\mathrm{f}})$ in the action \eqref{MSM2action4} defining our model. We now augment our model's action with the relevant constraints and carefully extract their consequences. 


Explicitly imposing these constraints 
in the action \eqref{MSM2action4} for Euclidean signature, we arrive at the augmented action
\begin{eqnarray}\label{MSM2action4constrained}
S_{\mathrm{cl}}[V_{2}]&=&\frac{\omega}{32\pi G}\int_{ t_{\mathrm{i}}}^{ t_{\mathrm{f}}}\mathrm{d} t\left[\frac{\dot{V}_{2}^{2}(t)}{\omega^{2}V_{2}(t)}-4\Lambda V_{2}(t)\right]+\lambda_{V_{3}}\left[\int_{ t_{\mathrm{i}}}^{ t_{\mathrm{f}}}\mathrm{d} t\,\omega V_{2}(t)-V_{3}\right]\nonumber\\ &&\qquad+\lambda_{\mathrm{i}}\left[\int_{ t_{\mathrm{i}}}^{ t_{\mathrm{f}}}\mathrm{d} t\,\omega\,\delta( t- t_{\mathrm{i}})V_{2}(t)-V_{2}( t_{\mathrm{i}})\right]+\lambda_{\mathrm{f}}\left[\int_{ t_{\mathrm{i}}}^{ t_{\mathrm{f}}}\mathrm{d} t\,\omega\,\delta( t- t_{\mathrm{f}})V_{2}(t)-V_{2}( t_{\mathrm{f}})\right]
\end{eqnarray}
in which $\lambda_{V_{3}}$ is the Lagrange multiplier associated with the constraint of fixed spacetime $3$-volume $V_{3}$, and $\lambda_{\mathrm{i}}$ and $\lambda_{\mathrm{f}}$ are the Lagrange multipliers associated with the constraints of fixed initial and final spatial $2$-volumes $V_{2}(t_{\mathrm{i}})$ and $V_{2}(t_{\mathrm{f}})$. The cosmological constant term also acts to constrain the spacetime $3$-volume $V_{3}$ with the cosmological constant itself serving as the associated Lagrange multiplier. We include the additional constraint of fixed $V_{3}$ to make our argument more transparent; in particular, we think of the cosmological constant $\Lambda$ as fixed and the Lagrange multiplier $\lambda_{V_{3}}$ as variable.

Varying the action \eqref{MSM2action4constrained} with respect to $V_{2}(t)$, we obtain the equation of motion 
\begin{equation}
2V_{2}(t)\ddot{V}_{2}(t)-\dot{V}_{2}^{2}(t)\pm4\omega^{2}(\Lambda-8\pi G\lambda_{V_{3}})V_{2}^{2}(t)=0,
\end{equation}
having the general solution
\begin{equation}\label{constrainedminisuperspacesolutions}
V_{2}( t)=\left\{\begin{array}{lcc} A\cos{[\omega\sqrt{\Lambda-8\pi G\lambda_{V_{3}}}(t-t_{0})]} & \mathrm{if} & \Lambda-8\pi G \lambda_{V_{3}}>0 \\ A\left(t-t_{0}\right)^{2} & \mathrm{if} & \Lambda-8\pi G\lambda_{V_{3}}=0 \\ A\cosh{[\omega\sqrt{8\pi G\lambda_{V_{3}}-\Lambda}(t-t_{0})]} & \mathrm{if} & \Lambda-8\pi G\lambda_{V_{3}}<0 \end{array}\right.
\end{equation}
for integration constants $A$ and $t_{0}$. Varying the action \eqref{MSM2action4constrained} with respect to $\lambda_{V_{3}}$, we obtain the constraint
\begin{equation}\label{V3constraint}
V_{3}=\int_{ t_{\mathrm{i}}}^{ t_{\mathrm{f}}}\mathrm{d} t\,\omega V_{2}( t),
\end{equation}
and varying the action \eqref{MSM2action4constrained} with respect to $\lambda_{\mathrm{i}}$ and $\lambda_{\mathrm{f}}$ constrains $V_{2}(t)$ to have the initial and final boundary values $V_{2}(t_{\mathrm{i}})$ and $V_{2}(t_{\mathrm{f}})$. 

We now focus on the spatial $2$-volume $V_{2}(t)$ for $\Lambda-8\pi G\lambda_{V_{3}}>0$ given in the first line of equation \eqref{constrainedminisuperspacesolutions}. Let $\ell_{\mathrm{eff}}^{-2}=\Lambda-8\pi G\lambda_{V_{3}}$. 
Recalling equation \eqref{dSvolprof}, we observe that
the spatial $2$-volume $V_{2}(t)$ for $\ell_{\mathrm{eff}}^{-2}>0$ is that of Euclidean de Sitter space if $A=4\pi\ell_{\mathrm{eff}}^{2}$. 
Assuming further that $V_{2}(t_{\mathrm{i}})=V_{2}(t_{\mathrm{f}})$ dictates that $t_{0}=0$. The difference $t_{\mathrm{f}}-t_{\mathrm{i}}$ (and, indeed, the value of $t_{\mathrm{i}}=-t_{\mathrm{f}}$) is then determined in terms of $V_{2}(t_{\mathrm{i}})=V_{2}(t_{\mathrm{f}})$ and $\ell_{\mathrm{eff}}$:
\begin{equation}\label{timeinterval}
\omega(t_{\mathrm{f}}-t_{\mathrm{i}})=\ell_{\mathrm{eff}}\cos^{-1}{\sqrt{\frac{V_{2}(t_{\mathrm{f}})}{4\pi\ell_{\mathrm{eff}}^{2}}}}.
\end{equation}
Substituting $V_{2}(t)$ for $\ell_{\mathrm{eff}}^{-2}>0$, $A=4\pi\ell_{\mathrm{eff}}^{2}$, and $t_{0}=0$ into equation \eqref{V3constraint}, we obtain
\begin{equation}\label{V3expression}
V_{3}=4\pi\ell_{\mathrm{eff}}^{3}\left[\cos^{-1}{\sqrt{\frac{V_{2}(t_{\mathrm{f}})}{4\pi\ell_{\mathrm{eff}}^{2}}}}+\sqrt{\frac{V_{2}(t_{\mathrm{f}})}{4\pi\ell_{\mathrm{eff}}^{2}}}\sqrt{1-\frac{V_{2}(t_{\mathrm{f}})}{4\pi\ell_{\mathrm{eff}}^{2}}}\right].
\end{equation}
Solving equation \eqref{V3expression} for $4\pi\ell_{\mathrm{eff}}^{2}$ and replacing $4\pi\ell_{\mathrm{eff}}^{2}$ in equation \eqref{constrainedminisuperspacesolutions}, we obtain
\begin{equation}\label{EdSvolprofconstrained}
V_{2}(t)=\frac{V_{3}}{\ell_{\mathrm{eff}}}\left[\cos^{-1}{\sqrt{\frac{V_{2}(t_{\mathrm{f}})}{4\pi\ell_{\mathrm{eff}}^{2}}}}+\sqrt{\frac{V_{2}(t_{\mathrm{f}})}{4\pi\ell_{\mathrm{eff}}^{2}}}\sqrt{1-\frac{V_{2}(t_{\mathrm{f}})}{4\pi\ell_{\mathrm{eff}}^{2}}}\right]^{-1}\cos^{2}{\left(\frac{\omega t}{\ell_{\mathrm{eff}}}\right)}.
\end{equation}
Equation \eqref{EdSvolprofconstrained} gives the spatial $2$-volume as a function of the global time coordinate of a portion of Euclidean de Sitter space constrained to have spacetime $3$-volume $V_{3}$ and initial and final boundary spatial $2$-volumes $V_{2}(t_{\mathrm{i}})=V_{2}(t_{\mathrm{f}})$. For given values of $G$ and $\Lambda$, with either the gauge fixing $\omega=\mathrm{constant}$ or the gauge fixing $t_{\mathrm{f}}=\mathrm{constant}$, we may choose values for $V_{3}$ and $V_{2}(t_{\mathrm{f}})$ and determine (if possible) the value of $\lambda_{V_{3}}$ dictated by the chosen values of $V_{3}$ and $V_{2}(t_{\mathrm{f}})$.
If $V_{2}( t_{\mathrm{f}})$ is not too large in comparison to $V_{3}$, then $\Lambda>8\pi G\lambda_{V_{3}}$, and the solution is a portion of Euclidean de Sitter space; however, if $V_{2}( t_{\mathrm{f}})$ is too large in comparison to $V_{3}$, then $\Lambda<8\pi G\lambda_{V_{3}}$, and the solution is a portion of Lorentzian de Sitter spacetime. 

\begin{figure}[!ht]
\centering
\setlength{\unitlength}{\textwidth}
\begin{picture}(1,0.3)
\put(0.0005,0.005){\includegraphics[scale=1]{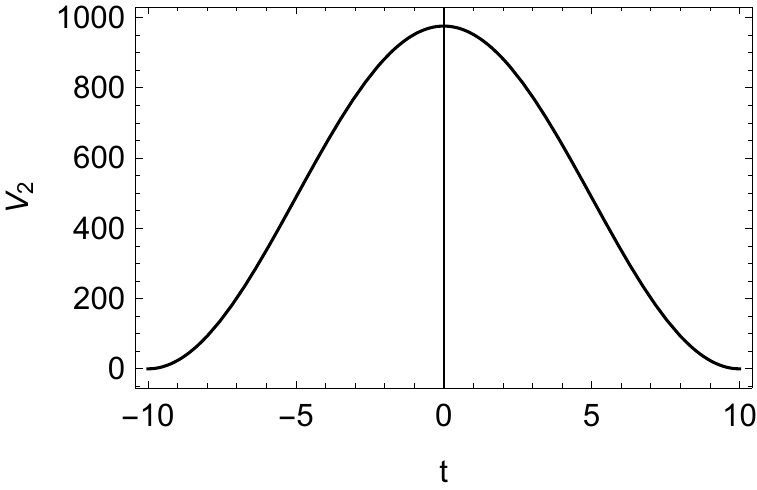}}
\put(0.51,0.00){\includegraphics[scale=1]{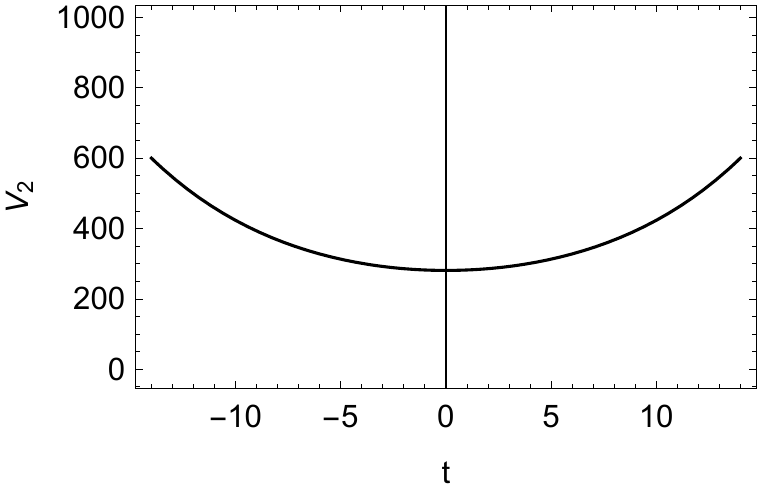}}
\put(0.26,-0.03){(a)}
\put(0.77,-0.03){(b)}
\end{picture}
\caption{Spatial $2$-volume $V_{2}(t)$ as a function of the global time coordinate $t$ for $A=4\pi\ell_{\mathrm{eff}}^{2}$, $t_{0}=0$, $G=1/8\pi$, and $\Lambda=1$ (a) $V_{3}=13500$, $V_{2}(t_{\mathrm{i}})=V_{2}(t_{\mathrm{f}})=0$, $\omega=1.38$, $\lambda_{V_{3}}=0.99$, and $\ell_{\mathrm{eff}}^{2}=77.63$ (b) $V_{3}=3300$, $V_{2}(t_{\mathrm{i}})=V_{2}(t_{\mathrm{f}})=600$, $\omega=0.31$, $\lambda_{V_{3}}=1.04$, and $\ell_{\mathrm{eff}}^{2}=-22.35$.}
\label{exampleV2}
\end{figure}

We now give examples of these two cases. Suppose that $G=1/8\pi$ and $\Lambda=1$. Choose first $V_{3}=13500$, $V_{2}( t_{\mathrm{f}})=0$, and $t_{\mathrm{f}}=10$. Equations \eqref{timeinterval} and \eqref{V3expression} then yield $\omega=1.38$ and $\lambda_{V_{3}}=0.99$ for which $\ell_{\mathrm{eff}}^{2}=77.63$. We display the spatial $2$-volume $V_{2}(t)$ for this case in figure \ref{exampleV2}(a). This first example models the circumstances of the Euclidean-like ensemble $\mathcal{E}_{\mathrm{E}}$: in this case $N_{2}^{\mathrm{SL}}(\tau_{\mathrm{f}})$ is not too large in comparison to $N_{3}$, so the discrete analogue $\mathcal{N}_{2}^{\mathrm{SL}}(\tau)$ of the spatial $2$-volume $V_{2}^{(\mathrm{EdS})}(t)$ of Euclidean de Sitter space accurately describes $\langle N_{2}^{\mathrm{SL}}(\tau)\rangle$. Compare figure \ref{exampleV2}(a) to figure \ref{volproffitT21V30K1IBFB4}.
Choose second $V_{3}=3300$, $V_{2}(t_{\mathrm{f}})=600$, and $t_{\mathrm{f}}=14$. Equations \eqref{timeinterval} and \eqref{V3expression} then yield $\omega=0.31$ and $\lambda_{V_{3}}=1.04$ for which $\ell_{\mathrm{eff}}^{2}=-22.35$. We display the spatial $2$-volume $V_{2}(t)$ for this case in figure \ref{exampleV2}(b). This second example models the circumstances of the Lorentzian-like ensemble $\mathcal{E}_{\mathrm{L}}$: in this case $N_{2}^{\mathrm{SL}}(\tau_{\mathrm{f}})$ is too large in comparison to $N_{3}$, so the discrete analogue $\mathcal{N}_{2}^{\mathrm{SL}}(\tau)$ of the spatial $2$-volume $V_{2}^{(\mathrm{LdS})}(t)$ of Lorentzian de Sitter spacetime accurately describes $\langle N_{2}^{\mathrm{SL}}(\tau)\rangle$. Compare figure \ref{exampleV2}(b) to figure \ref{nonminnonmin2fit}(a). The discrete analogue $\mathcal{N}_{2}^{\mathrm{SL}}(\tau)$ of the spatial $2$-volume $V_{2}^{(\mathrm{LdS})}(t)$ of Lorentzian de Sitter spacetime nevertheless arises from a model based on Euclidean Einstein gravity. Furthermore, the operator $\mathscr{M}(t,t')$ derived from the action \eqref{MSM2action4constrained} for linear perturbations $v_{2}(t)$ about the spatial $2$-volume $V_{2}(t)$ for $\ell_{\mathrm{eff}}^{2}<0$ coincides with the operator \eqref{vVMdLdS}, so the discrete analogue $\mathsf{n}_{2}^{\mathrm{SL}}(\tau)\,\mathsf{n}_{2}^{\mathrm{SL}}(\tau')$ of the connected $2$-point function $\mathbb{E}_{\mathrm{LdS}}[v_{2}(t)\,v_{2}(t')]$ still serves as the correct model for the connected $2$-point function $\langle n_{2}^{\mathrm{SL}}(\tau)\,n_{2}^{\mathrm{SL}}(\tau')\rangle$. 

The above discussion points towards an explanation of the measurements of $\langle N_{2}^{\mathrm{SL}}(\tau)\rangle$ and $\langle n_{2}^{\mathrm{SL}}(\tau)\,n_{2}^{\mathrm{SL}}(\tau')\rangle$ presented in section \ref{evidenceconjecture} and their analysis presented in section \ref{analysissupport} different from that of the conjecture of Cooperman and Miller. The model based on a minisuperspace truncation of Euclidean Einstein gravity also accurately describes $\langle N_{2}^{\mathrm{SL}}(\tau)\rangle$ and $\langle n_{2}^{\mathrm{SL}}(\tau)\,n_{2}^{\mathrm{SL}}(\tau')\rangle$ for the Lorentzian-like ensembles: the interaction of the constraints of fixed spacetime $3$-volume and fixed initial and final boundary spatial $2$-volumes forces $\langle N_{2}^{\mathrm{SL}}(\tau)\rangle$ and $\langle n_{2}^{\mathrm{SL}}(\tau)\,n_{2}^{\mathrm{SL}}(\tau')\rangle$ to be Lorentzian in form. 
We conclude accordingly that the geometries of causal triangulations comprising Lorentzian-like ensembles are not Lorentzian but Euclidean in nature. 

Our argument does not, however, clinch the case against the conjecture of Cooperman and Miller: had we run our reasoning starting from the action \eqref{MSM2action4} in Lorentzian signature, Euclidean de Sitter space would have arisen from Lorentzian de Sitter spacetime as the Lagrange multiplier $\lambda_{V_{3}}$ forced $\Lambda-8\pi G\lambda_{V_{3}}$ to change sign, and we would have concluded that geometries resembling Lorentzian de Sitter spacetime on sufficiently large scales dominate the ground state of causal dynamical triangulations. We chose to present our argument starting from the action \eqref{MSM2action4} in Euclidean signature because we know that the configurations simulated numerically must be Euclidean in nature: the Metropolis algorithm simply cannot handle complex contributions to the partition function \eqref{partitionfunctionfixedTN}. Still, we would like more definitive evidence for the Euclidean nature of the causal triangulations of Lorentzian-like ensembles represented in figures \ref{nonminnonminsame2} and \ref{nonminnonmin2}.

The two observables that we measured---$\langle N_{2}^{\mathrm{SL}}(\tau)\rangle$ and $\langle n_{2}^{\mathrm{SL}}(\tau)\,n_{2}^{\mathrm{SL}}(\tau')\rangle$---probe the quantum geometry defined by an ensemble of causal triangulations only on its largest scales. 
Since we do not consider observables that probe this quantum geometry on small scales, we do not assess the nature---Euclidean or Lorentzian---of the quantum geometry on small scales. To test the conjecture of Cooperman and Miller more definitively, we would like to make a statement regarding the nature of the quantum geometry on smaller scales, in particular, regarding the nature of local interactions, which should naively appear quite different if they are in fact Lorentzian. 
We should therefore probe the quantum geometry on small scales by measuring appropriate observables. 

\begin{figure}
\centering
\setlength{\unitlength}{\textwidth}
\begin{picture}(1,0.3)
\put(0.0005,0.005){\includegraphics[scale=1.5]{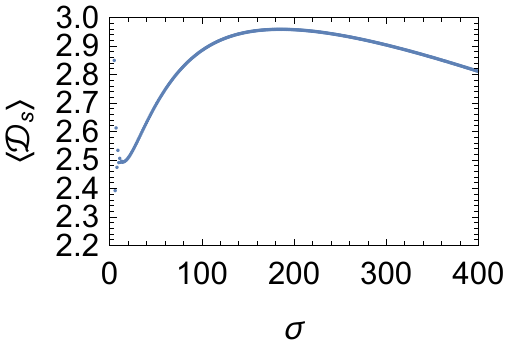}}
\put(0.51,0.00){\includegraphics[scale=1.5]{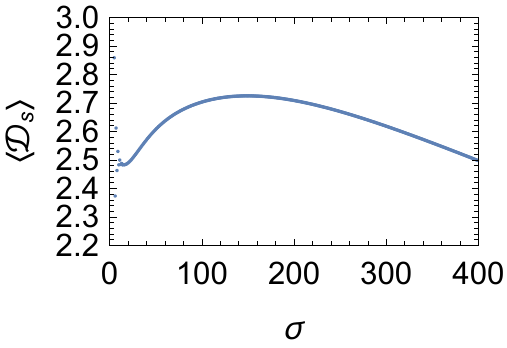}}
\put(0.26,-0.03){(a)}
\put(0.77,-0.03){(b)}
\end{picture}
\caption{Ensemble average spectral dimension $\langle\mathcal{D}_{\mathrm{s}}\rangle$ as a function of diffusion time $\sigma$ for $\bar{N}_{3}=30850$ and $k_{0}=1.00$ (a) $\bar{T}=21$ and $N_{2}^{\mathrm{SL}}(\mathsf{S}_{\mathrm{i}}^{2})=N_{2}^{\mathrm{SL}}(\mathsf{S}_{\mathrm{f}}^{2})=4$ (b) $\bar{T}=29$ and $N_{2}^{\mathrm{SL}}(\mathsf{S}_{\mathrm{i}}^{2})=N_{2}^{\mathrm{SL}}(\mathsf{S}_{\mathrm{f}}^{2})=600$.}
\label{specdimfig}
\end{figure}

Accordingly, we consider numerical measurements of the spectral dimension, a scale-dependent measure of the dimensionality of the quantum geometry, which probes the quantum geometry defined by an ensemble of causal triangulations on all scales. In appendix \ref{spectraldimension}, following several previous authors \cite{JA&JJ&RL6,JA&JJ&RL7,CA&SJC&JHC&PH&RKK&PZ,DB&JH,JHC,DNC&JJ,RK}, we define the spectral dimension $\mathcal{D}_{\mathrm{s}}(\sigma)$ as a function of the diffusion time $\sigma$, and we explain its numerical estimation. As in our analysis of the $2$-point function $\langle n_{2}^{\mathrm{SL}}(\tau)\,n_{2}^{\mathrm{SL}}(\tau')\rangle$, we compare the spectral dimension $\mathcal{D}_{\mathrm{s}}(\sigma)$ of the ensemble $\mathcal{E}_{\mathrm{E}}$ characterized by $\bar{T}=21$, $\bar{N}_{3}=30580$, $k_{0}=1.00$, and $N_{2}^{\mathrm{SL}}(\mathsf{S}_{\mathrm{i}}^{2})=N_{2}^{\mathrm{SL}}(\mathsf{S}_{\mathrm{f}}^{2})=4$ to the spectral dimension $\mathcal{D}_{\mathrm{s}}(\sigma)$ of the ensemble $\mathcal{E}_{\mathrm{L}}$ characterized by $\bar{T}=29$, $\bar{N}_{3}=30580$, $k_{0}=1.00$, and $N_{2}^{\mathrm{SL}}(\mathsf{S}_{\mathrm{i}}^{2})=N_{2}^{\mathrm{SL}}(\mathsf{S}_{\mathrm{f}}^{2})=600$. We display $\mathcal{D}_{\mathrm{s}}(\sigma)$ for the ensemble $\mathcal{E}_{\mathrm{E}}$ in figure \ref{specdimfig}(a) and for the ensemble $\mathcal{E}_{\mathrm{L}}$ in figure \ref{specdimfig}(b). 
The plot in figure \ref{specdimfig}(a) shows the behavior of $\mathcal{D}_{\mathrm{s}}(\sigma)$ previously understood as characteristic of phase C \cite{JA&JJ&RL6,JA&JJ&RL7,CA&SJC&JHC&PH&RKK&PZ,DB&JH,JHC,DNC&JJ,RK}. For intermediate diffusion times ($\sigma\sim200$ for $\mathcal{E}_{\mathrm{E}}$, $\sigma\sim150$ for $\mathcal{E}_{\mathrm{L}}$), the spectral dimension peaks at approximately the topological dimension of $3$; for smaller diffusion times ($\sigma\leq200$ for $\mathcal{E}_{\mathrm{E}}$, $\sigma\leq150$ for $\mathcal{E}_{\mathrm{L}}$), the spectral dimension dynamically reduces towards a value near $2$; and for larger diffusion times ($\sigma\geq200$ for $\mathcal{E}_{\mathrm{E}}$, $\sigma\geq150$ for $\mathcal{E}_{\mathrm{L}}$), the spectral dimension decays exponentially in the presence of positive curvature. The two measurements of $\mathcal{D}_{\mathrm{s}}(\sigma)$ displayed in figure \ref{specdimfig} exhibit essentially the same qualitative behavior and similar quantitative behavior. 
The maximal value of $\mathcal{D}_{\mathrm{s}}(\sigma)$ ($2.96$ for $\mathcal{E}_{\mathrm{E}}$, $2.72$ for $\mathcal{E}_{\mathrm{L}}$) is the primary difference. As Benedetti and Henson found for Euclidean-like ensembles \cite{DB&JH}, the depression of $\mathcal{D}_{\mathrm{s}}(\sigma)$ below the topological value of $3$ is a finite-size effect. We have verified that this depression 
is also a finite-size effect for Lorentzian-like ensembles. Although ensembles $\mathcal{E}_{\mathrm{E}}$ and $\mathcal{E}_{\mathrm{L}}$ are both characterized by $\bar{N}_{3}=30850$, we suspect that the ensemble $\mathcal{E}_{\mathrm{L}}$ exhibits stronger finite-size effects because the random walker can only probe a small portion of a quantum geometry resembling Lorentzian de Sitter spacetime on sufficiently large scales. Since $\mathcal{D}_{\mathrm{s}}(\sigma)$ for the ensemble $\mathcal{E}_{\mathrm{L}}$ behaves so similarly to $\mathcal{D}_{\mathrm{s}}(\sigma)$ for the ensemble $\mathcal{E}_{\mathrm{E}}$, we take these measurements of $\mathcal{D}_{\mathrm{s}}(\sigma)$ as evidence that the geometries of causal triangulations comprising the ensemble $\mathcal{E}_{\mathrm{L}}$ are Euclidean in nature, supporting our above conclusion.


\section{Lorentzian from Euclidean}\label{conclusion}

Studying the causal dynamical triangulations of $(2+1)$-dimensional Einstein gravity in the presence of initial and final spacelike boundaries, Cooperman and Miller identified several ensembles of causal triangulations the quantum geometry of which on sufficiently large scales appears to resemble closely that of Lorentzian de Sitter spacetime \cite{JHC&JMM}. On the basis of these findings, they conjectured that the partition function \eqref{partitionfunction} is dominated by causal triangulations the quantum geometry of which is nearly that of Lorentzian de Sitter spacetime on sufficiently large scales, possibly \emph{via} a mechanism akin to that of the Hartle-Hawking no-boundary proposal. The conjecture of Cooperman and Miller presented an exciting possibility: the definition of a Lorentzian quantum theory of gravity \emph{via} a Euclidean path integral, alleviating the necessity of reversing the Wick rotation of causal dynamical triangulations. We have argued for a much more plausible and mundane explanation of their findings: the implementation and interaction of multiple constraints may result in the partition function \eqref{partitionfunctionfixedTN} being dominated by (Euclidean) causal triangulations that closely resemble Lorentzian de Sitter spacetime on large scales. 
While not particularly exciting, our explanation adds one further piece of evidence for the proper behavior of the partition function defined \emph{via} causal dynamical triangulations. Our explanation also serves as a cautionary tale: beware hastily drawing conclusions regarding signs of signature change within the partition function \eqref{partitionfunction} of causal dynamical triangulations. 

The issue of reversing the Wick rotation of causal dynamical triangulations thus remains.
The results of modeling the large-scale quantum geometry within phase C on the basis of a minisuperspace truncation of Euclidean Einstein gravity, as exemplified by our modeling of the ensemble $\mathcal{E}_{\mathrm{E}}$ (and, indeed, also the ensemble $\mathcal{E}_{\mathrm{L}}$), suggest a straightforward possibility: 
since (Euclidean) causal triangulations resembling Euclidean de Sitter space on sufficiently large scales dominate the partition function \eqref{partitionfunction}, obtained by Wick rotation from the path sum \eqref{causalpathsum}, (Lorentzian) causal triangulations resembling Lorentzian de Sitter spacetime on sufficiently large scales dominate the path sum \eqref{causalpathsum}. 
For this interpretation to have force, one must establish a rigorous path from the Euclidean theory to the Lorentzian theory by demonstrating an Osterwalder-Schrader-type theorem for causal dynamical triangulations. Although technically challenging, achieving such a theorem is likely within reach since the action $\mathcal{S}_{\mathrm{cl}}^{(\mathrm{E})}(\mathcal{T}_{c})$ for Einstein gravity is reflection-positive, a key axiom of the Osterwalder-Schrader reconstruction theorem. We maintain that the promising results of causal dynamical triangulations warrant such an effort.

\section*{Acknowledgments}

We thank Christian Anderson, David Kamensky, and especially Rajesh Kommu for allowing us to employ parts of their computer codes. We also thank Joe Henson, Ian Morrison, Erik Schnetter, and especially Steve Carlip and Renate Loll for useful discussions. JHC acknowledges support from the Department of Energy under grant DE-FG02-91ER40674 at the University of California, Davis and from Stichting voor Fundamenteel Onderzoek der Materie itself supported by Nederlandse Organisatie voor Wetenschappelijk Onderzoek. KL acknowledges support from the SURF program at Chapman University and the hospitality of the Department of Physics at the University of California, Davis. JMM acknowledge support from the National Science Foundation under REU grant PHY-1004848 at the University of California, Davis. This work utilized the Janus supercomputer, which is supported by the National Science Foundation (award number CNS-0821794) and the University of Colorado, Boulder. The Janus supercomputer is a joint effort of the University of Colorado, Boulder, the University of Colorado, Denver, and the National Center for Atmospheric Research. This research was supported in part by the Perimeter Institute for Theoretical Physics. Research at the Perimeter Institute is supported by the Government of Canada through Industry Canada and by the Province of Ontario through the Ministry of Economic Development and Innovation.

\appendix

\section{Action $\mathcal{S}_{\mathrm{cl}}^{(\mathrm{E})}[\mathcal{T}_{c}]$}\label{completediscreteEaction}

Cooperman and Miller constructed the action $\mathcal{S}_{\mathrm{cl}}^{(\mathrm{E})}[\mathcal{T}_{c}]$ in the case of $(2+1)$-dimensional Einstein gravity for spacetime topology of the direct product of a $2$-sphere $\mathsf{S}^{2}$ and a real interval $\mathsf{I}$, finding that
\begin{eqnarray}\label{completeEaction}
\mathcal{S}_{\mathrm{cl}}^{(\mathrm{E})}[\mathcal{T}_{c}]&=&\frac{ia}{8\pi G}\left[\frac{2\pi}{i}\left(N_{1}^{\mathrm{SL}}-N_{1}^{\mathrm{SL}}(\mathsf{S}_{\mathrm{i}}^{2})-N_{1}^{\mathrm{SL}}(\mathsf{S}_{\mathrm{f}}^{2})\right)-\frac{1}{i}\vartheta_{\mathrm{SL}}^{(2,2)}\left(2N_{3}^{(2,2)}-N_{3\uparrow}^{(2,2)}(\mathsf{S}_{\mathrm{i}}^{2})-N_{3\downarrow}^{(2,2)}(\mathsf{S}_{\mathrm{f}}^{2})\right)\right.\nonumber\\ && \qquad\qquad\left.-\frac{1}{i}\vartheta_{\mathrm{SL}}^{(1,3)}\left(4N_{1}^{\mathrm{SL}}-2N_{1}^{\mathrm{SL}}(\mathsf{S}_{\mathrm{i}}^{2})-2N_{1}^{\mathrm{SL}}(\mathsf{S}_{\mathrm{f}}^{2})\right)-2\pi i\sqrt{-\alpha}N_{1}^{\mathrm{TL}}+4i\sqrt{-\alpha}\vartheta_{\mathrm{TL}}^{(2,2)}N_{3}^{(2,2)}\right.\nonumber\\ &&\left.\qquad\qquad+3i\sqrt{-\alpha}\vartheta_{\mathrm{TL}}^{(1,3)}N_{3}^{(1,3)}+3i\sqrt{-\alpha}\vartheta_{\mathrm{TL}}^{(3,1)}N_{3}^{(3,1)}\right]\nonumber\\ && -\frac{i\Lambda}{8\pi G}\left[\mathcal{V}_{3}^{(2,2)}N_{3}^{(2,2)}+\mathcal{V}_{3}^{(1,3)}N_{3}^{(1,3)}+\mathcal{V}_{3}^{(3,1)}N_{3}^{(3,1)}\right]\nonumber\\ && +\frac{ia}{8\pi G}\left[\frac{\pi}{i} N_{1}^{\mathrm{SL}}(\mathsf{S}_{\mathrm{i}}^{2})-\frac{2}{i}\vartheta_{\mathrm{SL}}^{(3,1)}N_{1}^{\mathrm{SL}}(\mathsf{S}_{\mathrm{i}}^{2})-\frac{1}{i}\vartheta_{\mathrm{SL}}^{(2,2)}N_{3\uparrow}^{(2,2)}(\mathsf{S}_{\mathrm{i}}^{2})\right]\nonumber\\ &&+\frac{ia}{8\pi G}\left[\frac{\pi}{i} N_{1}^{\mathrm{SL}}(\mathsf{S}_{\mathrm{f}}^{2})-\frac{2}{i}\vartheta_{\mathrm{SL}}^{(3,1)}N_{1}^{\mathrm{SL}}(\mathsf{S}_{\mathrm{f}}^{2})-\frac{1}{i}\vartheta_{\mathrm{SL}}^{(2,2)}N_{3\downarrow}^{(2,2)}(\mathsf{S}_{\mathrm{f}}^{2})\right].
\end{eqnarray}
We refer the reader to \cite{JHC&JMM} for the derivation of the action \eqref{completeEaction}. $N_{1}^{\mathrm{SL}}$ is the number of spacelike $1$-simplices (edges) and $N_{1}^{\mathrm{TL}}$ is the number of timelike $1$-simplices. $N_{3\uparrow}^{(2,2)}(\mathsf{S}^{2})$ is the number of future-directed $(2,2)$ $3$-simplices attached to the $2$-sphere $\mathsf{S}^{2}$, and $N_{3\downarrow}^{(2,2)}(\mathsf{S}^{2})$ is the number of past-directed $(2,2)$ $3$-simplices attached to the $2$-sphere $\mathsf{S}^{2}$. $\vartheta_{\mathrm{SL}}^{(p,q)}$ is the Euclidean dihedral angle about a spacelike $1$-simplex of a $(p,q)$ $3$-simplex, $\vartheta_{\mathrm{TL}}^{(p,q)}$ is the Euclidean dihedral angle about a timelike $1$-simplex of a $(p,q)$ $3$-simplex, and $\mathcal{V}_{3}^{(p,q)}$ is the Euclidean spacetime $3$-volume of a $(p,q)$ $3$-simplex. We refer the reader to \cite{JA&JJ&RL2,CA&SJC&JHC&PH&RKK&PZ} for explicit expressions for these Euclidean dihedral angles and spacetime $3$-volumes.

\section{Derivation of $\mathcal{N}_{2}^{\mathrm{SL}}(\tau)$}\label{derivation1pt}

We derive the discrete analogue $\mathcal{N}_{2}^{\mathrm{SL}}(\tau)$ of the spatial $2$-volume $V_{2}^{(\mathrm{EdS})}( t)$ as a function of the global time coordinate $t$ of Euclidean de Sitter space, given in equation \eqref{dSvolprof}, and of the spatial $2$-volume $V_{2}^{(\mathrm{LdS})}(t)$ as a function of the global time coordinate $ t$ of Lorentzian de Sitter spacetime, given in equation \eqref{LdSvolprof}. 



We start from the doubling scaling limit of the discrete spacetime $3$-volume given in equation \eqref{FSSansatz}:
\begin{equation}\label{naiveCL}
V_{3}=\lim_{\substack{N_{3}\rightarrow\infty \\ a\rightarrow0}}C_{3}N_{3}a^{3}.
\end{equation}
Assuming that equation \eqref{naiveCL} holds for finite number $N_{3}$ of $3$-simplices and lattice spacing $a$ without significant corrections, we express equation \eqref{naiveCL} with its left hand side as an integral over the global time coordinate $ t$ and its right hand side as a sum over the discrete time coordinate $\tau$:
\begin{equation}\label{integratednaiveCL}
\int_{t_{\mathrm{i}}}^{ t_{\mathrm{f}}}\mathrm{d} t\,\omega V_{2}( t)=2C_{3}a^{3}(1+\xi)\sum_{\tau=1}^{\bar{T}}N_{2}^{\mathrm{SL}}(\tau).
\end{equation}
According to the finite-size scaling \emph{Ansatz} based on equation \eqref{naiveCL}, in the combination of the infinite volume and continuum limits, we expect the relation 
\begin{equation}\label{timescaling}
\frac{\tau}{N_{3}^{1/3}}=\frac{ t}{V_{3}^{1/3}}
\end{equation}
between $\tau$ and $ t$ and the relation
\begin{equation}\label{spacescaling}
\frac{N_{2}^{\mathrm{SL}}}{N_{3}^{2/3}}=\frac{V_{2}}{V_{3}^{2/3}}
\end{equation}
between $N_{2}^{\mathrm{SL}}$ and $V_{2}$. Constants of proportionality in the relations \eqref{timescaling} and \eqref{spacescaling} are redundant for the following derivation.
In these limits we identify the integral $\int\mathrm{d} t\,V_{3}^{-1/3}$ with the sum $\sum_{\tau}\Delta\tau\,N_{3}^{-1/3}$ in equation \eqref{integratednaiveCL}, yielding
\begin{equation}
\omega V_{3}^{1/3}V_{2}( t)=2C_{3}a^{3}(1+\xi)N_{3}^{1/3}N_{2}^{\mathrm{SL}}(\tau).
\end{equation}
Solving for $N_{2}^{\mathrm{SL}}(\tau)$, we obtain
\begin{equation}\label{N2V2}
N_{2}^{\mathrm{SL}}(\tau)=\frac{\omega V_{3}^{1/3}}{2C_{3}N_{3}^{1/3}a^{3}(1+\xi)}V_{2}( t).
\end{equation}
We next need to substitute appropriate expressions for the spatial $2$-volume $V_{2}(t)$. Considering the measurements of ensemble average $\langle N_{2}^{\mathrm{SL}}(\tau)\rangle$ presented in section \ref{evidenceconjecture}, we consider the finite portions of Euclidean de Sitter space and of Lorentzian de Sitter spacetime for which $ t\in[t_{\mathrm{i}},t_{\mathrm{f}}]$. 
This portion of Euclidean de Sitter space has spacetime $3$-volume
\begin{equation}\label{V3EdS}
V_{3}=\int_{t_{\mathrm{i}}}^{ t_{\mathrm{f}}}\mathrm{d} t\,\omega V_{2}^{(\mathrm{EdS})}( t)=2\pi\ell_{\mathrm{dS}}^{3}\left\{\frac{\omega(t_{\mathrm{f}}-t_{\mathrm{i}})}{\ell_{\mathrm{dS}}}+\sin{\left[\frac{\omega(t_{\mathrm{f}}-t_{\mathrm{i}})}{\ell_{\mathrm{dS}}}\right]}\cos{\left[\frac{\omega(t_{\mathrm{f}}+t_{\mathrm{i}})}{\ell_{\mathrm{dS}}}\right]}\right\}, 
\end{equation}
which is equivalent to equation \eqref{V3expression}, and this portion of Lorentzian de Sitter spacetime has spacetime $3$-volume
\begin{equation}\label{V3LdS}
V_{3}=\int_{t_{\mathrm{i}}}^{ t_{\mathrm{f}}}\mathrm{d} t\,\omega V_{2}^{(\mathrm{LdS})}( t)=2\pi\ell_{\mathrm{dS}}^{3}\left\{\frac{\omega ( t_{\mathrm{f}}-t_{\mathrm{i}})}{\ell_{\mathrm{dS}}}+\sinh{\left[\frac{\omega(t_{\mathrm{f}}-t_{\mathrm{i}})}{\ell_{\mathrm{dS}}}\right]}\cosh{\left[\frac{\omega ( t_{\mathrm{f}}+t_{\mathrm{i}})}{\ell_{\mathrm{dS}}}\right]}\right\}.
\end{equation}
Solving equation \eqref{V3EdS} for $4\pi\ell_{\mathrm{dS}}^{2}$ in terms of $V_{3}$ and substituting into equation \eqref{dSvolprof} yields
\begin{equation}\label{dSvolprofV3}
V_{2}^{(\mathrm{EdS})}(t)=\frac{2V_{3}}{\ell_{\mathrm{dS}}}\left\{\frac{\omega(t_{\mathrm{f}}-t_{\mathrm{i}})}{\ell_{\mathrm{dS}}}+\sin{\left[\frac{\omega(t_{\mathrm{f}}-t_{\mathrm{i}})}{\ell_{\mathrm{dS}}}\right]}\cos{\left[\frac{\omega(t_{\mathrm{f}}+t_{\mathrm{i}})}{\ell_{\mathrm{dS}}}\right]}\right\}^{-1}\cos^{2}{\left(\frac{\omega t}{\ell_{\mathrm{dS}}}\right)},
\end{equation}
while solving equation \eqref{V3LdS} for $4\pi\ell_{\mathrm{dS}}^{2}$ in terms of $V_{3}$ and substituting into equation \eqref{LdSvolprof} yields
\begin{equation}\label{LdSvolprofV3}
V_{2}^{(\mathrm{LdS})}( t)=\frac{2V_{3}}{\ell_{\mathrm{dS}}}\left\{\frac{\omega ( t_{\mathrm{f}}-t_{\mathrm{i}})}{\ell_{\mathrm{dS}}}+\sinh{\left[\frac{\omega(t_{\mathrm{f}}-t_{\mathrm{i}})}{\ell_{\mathrm{dS}}}\right]}\cosh{\left[\frac{\omega ( t_{\mathrm{f}}+t_{\mathrm{i}})}{\ell_{\mathrm{dS}}}\right]}\right\}^{-1}\cosh^{2}{\left(\frac{\omega  t}{\ell_{\mathrm{dS}}}\right)}.
\end{equation}
We substitute equation \eqref{dSvolprofV3} into equation \eqref{N2V2}, obtaining
\begin{equation}\label{N2V2EdS}
N_{2}^{\mathrm{SL}}(\tau)=\frac{\omega V_{3}^{1/3}}{2C_{3}N_{3}^{1/3}a^{3}(1+\xi)}\frac{2V_{3}}{\ell_{\mathrm{dS}}}\left\{\frac{\omega(t_{\mathrm{f}}-t_{\mathrm{i}})}{\ell_{\mathrm{dS}}}+\sin{\left[\frac{\omega(t_{\mathrm{f}}-t_{\mathrm{i}})}{\ell_{\mathrm{dS}}}\right]}\cos{\left[\frac{\omega(t_{\mathrm{f}}+t_{\mathrm{i}})}{\ell_{\mathrm{dS}}}\right]}\right\}^{-1}\cos^{2}{\left(\frac{\omega t}{\ell_{\mathrm{dS}}}\right)},
\end{equation}
and we substitute equation \eqref{LdSvolprofV3} into equation \eqref{N2V2}, obtaining
\begin{equation}\label{N2V2LdS}
N_{2}^{\mathrm{SL}}(\tau)=\frac{\omega V_{3}^{1/3}}{2C_{3}N_{3}^{1/3}a^{3}(1+\xi)}\frac{2V_{3}}{\ell_{\mathrm{dS}}}\left\{\frac{\omega ( t_{\mathrm{f}}-t_{\mathrm{i}})}{\ell_{\mathrm{dS}}}+\sinh{\left[\frac{\omega(t_{\mathrm{f}}-t_{\mathrm{i}})}{\ell_{\mathrm{dS}}}\right]}\cosh{\left[\frac{\omega ( t_{\mathrm{f}}+t_{\mathrm{i}})}{\ell_{\mathrm{dS}}}\right]}\right\}^{-1}\cosh^{2}{\left(\frac{\omega  t}{\ell_{\mathrm{dS}}}\right)}.
\end{equation}
Using equation \eqref{naiveCL} and replacing $t$ with $V_{3}^{1/3}\tau/N_{3}^{1/3}$ according to relation \eqref{timescaling}, equation \eqref{N2V2EdS} becomes
\begin{equation}
N_{2}^{\mathrm{SL}}(\tau)=\frac{\omega V_{3}^{1/3}N_{3}\cos^{2}{\left(\frac{\omega V_{3}^{1/3}\tau}{\ell_{\mathrm{dS}}N_{3}^{1/3}}\right)}}{N_{3}^{1/3}(1+\xi)\ell_{\mathrm{dS}}\left\{\frac{\omega V_{3}^{1/3}(\tau_{\mathrm{f}}-\tau_{\mathrm{i}})}{\ell_{\mathrm{dS}}N_{3}^{1/3}}+\sin{\left[\frac{\omega V_{3}^{1/3}(\tau_{\mathrm{f}}-\tau_{\mathrm{i}})}{\ell_{\mathrm{dS}}N_{3}^{1/3}}\right]}\cos{\left[\frac{\omega V_{3}^{1/3}(\tau_{\mathrm{f}}+\tau_{\mathrm{i}})}{\ell_{\mathrm{dS}}N_{3}^{1/3}}\right]}\right\}}.
\end{equation}
and equation \eqref{N2V2LdS} becomes
\begin{equation}
N_{2}^{\mathrm{SL}}(\tau)=\frac{\omega V_{3}^{1/3}N_{3}\cosh^{2}{\left(\frac{\omega V_{3}^{1/3}\tau}{\ell_{\mathrm{dS}}N_{3}^{1/3}}\right)}}{N_{3}^{1/3}(1+\xi)\ell_{\mathrm{dS}}\left\{\frac{\omega V_{3}^{1/3}(\tau_{\mathrm{f}}-\tau_{\mathrm{i}})}{\ell_{\mathrm{dS}}N_{3}^{1/3}}+\sinh{\left[\frac{\omega V_{3}^{1/3}(\tau_{\mathrm{f}}-\tau_{\mathrm{i}})}{\ell_{\mathrm{dS}}N_{3}^{1/3}}\right]}\cosh{\left[\frac{\omega V_{3}^{1/3}(\tau_{\mathrm{f}}+\tau_{\mathrm{i}})}{\ell_{\mathrm{dS}}N_{3}^{1/3}}\right]}\right\}}.
\end{equation}
Substituting $N_{3}^{(1,3)}$ for $N_{3}$ according to the identity $N_{3}=2(1+\xi)N_{3}^{(1,3)}$ and defining the parameter
\begin{equation}\label{s0}
\bar{s}_{0}=\frac{2^{1/3}(1+\xi)^{1/3}\ell_{\mathrm{dS}}}{\omega V_{3}^{1/3}},
\end{equation}
we finally arrive at the discrete analogue $\mathcal{N}_{2}^{\mathrm{SL}}(\tau)$ of the spatial $2$-volume $V_{2}^{(\mathrm{EdS})}(t)$, 
\begin{equation}\label{discreteanalogueEdS}
\mathcal{N}_{2}^{\mathrm{SL}}(\tau)=\frac{\langle N_{3}^{(1,3)}\rangle\cos^{2}{\left(\frac{\tau}{\bar{s}_{0}\langle N_{3}^{(1,3)}\rangle^{1/3}}\right)}}{\bar{s}_{0}\langle N_{3}^{(1,3)}\rangle^{1/3}\left\{\frac{(\tau_{\mathrm{f}}+\tau_{\mathrm{i}})}{\bar{s}_{0}\langle N_{3}^{(1,3)}\rangle^{1/3}}+\sin{\left[\frac{(\tau_{\mathrm{f}}-\tau_{\mathrm{i}})}{\bar{s}_{0}\langle N_{3}^{(1,3)}\rangle^{1/3}}\right]}\cos{\left[\frac{(\tau_{\mathrm{f}}+\tau_{\mathrm{i}})}{\bar{s}_{0}\langle N_{3}^{(1,3)}\rangle^{1/3}}\right]}\right\}}.
\end{equation}
and the discrete analogue $\mathcal{N}_{2}^{\mathrm{SL}}(\tau)$ of the spatial $2$-volume $V_{2}^{(\mathrm{LdS})}(t)$, 
\begin{equation}\label{discreteanalogueLdS}
\mathcal{N}_{2}^{\mathrm{SL}}(\tau)=\frac{\langle N_{3}^{(1,3)}\rangle\cosh^{2}{\left(\frac{\tau}{\bar{s}_{0}\langle N_{3}^{(1,3)}\rangle^{1/3}}\right)}}{\bar{s}_{0}\langle N_{3}^{(1,3)}\rangle^{1/3}\left\{\frac{(\tau_{\mathrm{f}}+\tau_{\mathrm{i}})}{\bar{s}_{0}\langle N_{3}^{(1,3)}\rangle^{1/3}}+\sinh{\left[\frac{(\tau_{\mathrm{f}}-\tau_{\mathrm{i}})}{\bar{s}_{0}\langle N_{3}^{(1,3)}\rangle^{1/3}}\right]}\cosh{\left[\frac{(\tau_{\mathrm{f}}+\tau_{\mathrm{i}})}{\bar{s}_{0}\langle N_{3}^{(1,3)}\rangle^{1/3}}\right]}\right\}}.
\end{equation}
For the case in which $\tau_{\mathrm{i}}=-\bar{T}/2$ and $\tau_{\mathrm{f}}=\bar{T}/2$, equation \eqref{discreteanalogueEdS} simplifies to 
\begin{equation}
\mathcal{N}_{2}^{\mathrm{SL}}(\tau)=\frac{\langle N_{3}^{(1,3)}\rangle}{\bar{s}_{0}\langle N_{3}^{(1,3)}\rangle^{1/3}}\left[\frac{\bar{T}}{\bar{s}_{0}\langle N_{3}^{(1,3)}\rangle^{1/3}}+\sin{\left(\frac{\bar{T}}{\bar{s}_{0}\langle N_{3}^{(1,3)}\rangle^{1/3}}\right)}\right]^{-1}\cos^{2}{\left(\frac{\tau}{\bar{s}_{0}\langle N_{3}^{(1,3)}\rangle^{1/3}}\right)},
\end{equation}
and equation \eqref{discreteanalogueLdS} simplifies to
\begin{equation}
\mathcal{N}_{2}^{\mathrm{SL}}(\tau)=\frac{\langle N_{3}^{(1,3)}\rangle}{\bar{s}_{0}\langle N_{3}^{(1,3)}\rangle^{1/3}}\left[\frac{\bar{T}}{\bar{s}_{0}\langle N_{3}^{(1,3)}\rangle^{1/3}}+\sinh{\left(\frac{\bar{T}}{\bar{s}_{0}\langle N_{3}^{(1,3)}\rangle^{1/3}}\right)}\right]^{-1}\cosh^{2}{\left(\frac{\tau}{\bar{s}_{0}\langle N_{3}^{(1,3)}\rangle^{1/3}}\right)}.
\end{equation}

\section{Derivation of $\mathsf{n}_{2}^{\mathrm{SL}}(\tau)\,\mathsf{n}_{2}^{\mathrm{SL}}(\tau')$}\label{derivation2pt}

We derive the discrete analogue $\mathsf{n}_{2}^{\mathrm{SL}}(\tau)\,\mathsf{n}_{2}^{\mathrm{SL}}(\tau')$ of the connected $2$-point function $\mathbb{E}_{\mathrm{EdS}}[v_{2}( t)\,v_{2}( t)]$ of gravitational perturbations $v_{2}(t)$ of the spatial $2$-volume $V_{2}^{(\mathrm{EdS})}(t)$ of Euclidean de Sitter space and of the connected $2$-point function $\mathbb{E}_{\mathrm{LdS}}[v_{2}( t)\,v_{2}( t)]$ of gravitational perturbations $v_{2}(t)$ of the spatial $2$-volume $V_{2}^{(\mathrm{LdS})}(t)$ of Lorentzian de Sitter spacetime.
We start from equations \eqref{vVMdEdS} and \eqref{vVMdLdS}, the expressions for the van Vleck-Morette determinants $\mathscr{M}(t,t')$. We first discretize the operator $\mathscr{M}(t,t')$ 
on a $1$-dimensional lattice of $\bar{T}$ sites, transforming the differential operators of equations \eqref{vVMdEdS} and \eqref{vVMdLdS} into finite-difference operators. Specifically, replacing $t$ with $V_{3}^{1/3}\tau/N_{3}^{1/3}$ according to relation \eqref{timescaling}, substituting $N_{3}^{(1,3)}$ for $N_{3}$ according to the identity $N_{3}=2(1+\xi)N_{3}^{(1,3)}$, and employing the definition \eqref{s0} of the fit parameter $\bar{s}_{0}$, equation \eqref{vVMdEdS} becomes
\begin{eqnarray}
\mathsf{M}(\tau,\tau')&=&\frac{1}{64\pi^{2}G\ell_{\mathrm{dS}}^{3}}\sec^{2}{\left(\frac{\tau}{\bar{s}_{0}\langle N_{3}^{(1,3)}\rangle^{1/3}}\right)}\Bigg[\frac{\Delta^{2}}{\Delta\tau^{2}}\nonumber\\ && \qquad\qquad\qquad\qquad\qquad+2\tan{\left(\frac{\tau}{\bar{s}_{0}\langle N_{3}^{(1,3)}\rangle^{1/3}}\right)}\frac{\Delta}{\Delta\tau}+2\sec^{2}{\left(\frac{\tau}{\bar{s}_{0}\langle N_{3}^{(1,3)}\rangle^{1/3}}\right)}\Bigg],
\end{eqnarray}
and equation \eqref{vVMdLdS} becomes
\begin{eqnarray}
\mathsf{M}(\tau,\tau')&=&\frac{1}{64\pi^{2}G\ell_{\mathrm{dS}}^{3}}\sech^{2}{\left(\frac{\tau}{\bar{s}_{0}\langle N_{3}^{(1,3)}\rangle^{1/3}}\right)}\Bigg[\frac{\Delta^{2}}{\Delta\tau^{2}}\nonumber\\ && \qquad\qquad\qquad\qquad\qquad-2\tanh{\left(\frac{\tau}{\bar{s}_{0}\langle N_{3}^{(1,3)}\rangle^{1/3}}\right)}\frac{\Delta}{\Delta\tau}-2\sech^{2}{\left(\frac{\tau}{\bar{s}_{0}\langle N_{3}^{(1,3)}\rangle^{1/3}}\right)}\Bigg].
\end{eqnarray}
$\Delta/\Delta\tau$ and $\Delta^{2}/\Delta\tau^{2}$ denote appropriate finite-difference operators. $\mathsf{M}(\tau,\tau')$ is now just a $\bar{T}\times\bar{T}$ symmetric matrix. We next add to $\mathsf{M}(\tau,\tau')$ two $\bar{T}\times\bar{T}$ matrices: one implementing the constraint \eqref{eigenfunctionconstraint} and one enforcing the boundary conditions $\nu_{j}( t_{\mathrm{i}})=0$ and $\nu_{j}( t_{\mathrm{f}})=0$. 
We finally numerically diagonalize the constrained operator $\mathsf{M}(\tau,\tau')$ to obtain its eigenvectors $\nu_{j}(\tau)$ and associated eigenvalues $\mu_{j}$. We input the value of $\bar{s}_{0}$ obtained from the best fit of $\mathcal{N}_{2}^{\mathrm{SL}}(\tau)$ to $\langle N_{2}^{\mathrm{SL}}(\tau)\rangle$, and we scale each $\mu_{j}$ by an overall constant, corresponding to the value of the coefficient $1/64\pi^{2}\hbar G \ell_{\mathrm{dS}}^{3}$, 
obtained by exactly matching the values of $\mu_{1}$ and $\lambda_{1}$. 

\section{Definition and measurement of the spectral dimension}\label{spectraldimension}

The spectral dimension, a measure of the dimensionality of a space as experienced by a diffusing random walker, is defined \emph{via} the heat equation governing this walker's diffusion. On a Wick-rotated causal triangulation the integrated heat equation takes the form
\begin{eqnarray}\label{heatequation}
\mathcal{K}_{\mathcal{T}_{c}}(s,s',\sigma+1)=(1-\varrho)\mathcal{K}_{\mathcal{T}_{c}}(s,s',\sigma)+\frac{\varrho}{N(\mathscr{N}(s))}\sum_{s''\in\mathscr{N}(s)}\mathcal{K}_{\mathcal{T}_{c}}(s'',s',\sigma).
\end{eqnarray}
The heat kernel $\mathcal{K}_{\mathcal{T}_{c}}(s,s',\sigma)$ gives the probability of diffusion from $D$-simplex $s$ to $D$-simplex $s'$ (or \emph{vice versa}) in $\sigma$ diffusion time steps; the diffusion constant $\varrho$ characterizes the dwell probability in a given time step; and $\mathscr{N}(s)$ is the set of $N(\mathscr{N}(s))$ nearest neighbors of the $D$-simplex $s$. We set $\varrho=4/5$. 
The heat trace or return probability, defined as 
\begin{equation}
\mathcal{P}_{\mathcal{T}_{c}}(\sigma)=\frac{1}{N_{D}}\sum_{s\in\mathcal{T}_{c}}\mathcal{K}_{\mathcal{T}_{c}}(s,s,\sigma),
\end{equation}
gives the probability for a random walker to return to its starting $D$-simplex in $\sigma$ diffusion time steps. The spectral dimension $\mathcal{D}_{\mathrm{s}}^{(\mathcal{T}_{c})}(\sigma)$ quantifies the scaling of the return probability $\mathcal{P}_{\mathcal{T}_{c}}(\sigma)$ with diffusion time $\sigma$:
\begin{equation}\label{specdim}
\mathcal{D}_{\mathrm{s}}^{(\mathcal{T}_{c})}(\sigma)=-2\frac{\mathrm{d}\ln{\mathcal{P}_{\mathcal{T}_{c}}(\sigma)}}{\mathrm{d}\ln{\sigma}}.
\end{equation}
The definition \eqref{specdim} is primarily motivated by the fact that, for diffusion of a random walker on a continuous Riemannian manifold, the spectral dimension at zero diffusion time coincides with this manifold's topological dimension.

Given an ensemble of causal triangulations representative of those contributing to the partition function \eqref{partitionfunctionfixedTN}, we numerically estimate the spectral dimension $\mathcal{D}_{\mathrm{s}}(\sigma)$ as follows. The number $N_{D}$ of $D$-simplices comprising a typical causal triangulation $\mathcal{T}_{c}$ is of order $10^{5}$, so we estimate the return probability $\mathcal{P}_{\mathcal{T}_{c}}(\sigma)$ by considering only a subset of $K$ randomly selected $D$-simplices $s_{k}$:
\begin{equation}
\mathcal{P}_{\mathcal{T}_{c}}^{(K)}(\sigma)=\frac{1}{K}\sum_{s_{k}\in\mathcal{T}_{c}}\mathcal{K}_{\mathcal{T}_{c}}(s_{k},s_{k},\sigma).
\end{equation}
One clearly recovers the return probability $\mathcal{P}_{\mathcal{T}_{c}}(\sigma)$ in the limit as $K$ approaches $N_{D}$:
\begin{equation}
\mathcal{P}_{\mathcal{T}_{c}}(\sigma)=\lim_{K\rightarrow N_{D}}\mathcal{P}_{\mathcal{T}_{c}}^{(K)}(\sigma).
\end{equation}
Since the number $N(\mathcal{T}_{c})$ of causal triangulations comprising an ensemble is necessarily finite, we estimate the expectation value $\mathbb{E}[\mathcal{P}(\sigma)]$ of the return probability $\mathcal{P}(\sigma)$ by its average over an ensemble:
\begin{equation}
\langle\mathcal{P}(\sigma)\rangle=\frac{1}{N(\mathcal{T}_{c})}\sum_{j=1}^{N(\mathcal{T}_{c})}\mathcal{P}_{\mathcal{T}_{c}^{(j)}}(\sigma).
\end{equation}
One clearly recovers the expectation value $\mathbb{E}[\mathcal{P}(\sigma)]$ in the limit as $N(\mathcal{T}_{c})$ diverges without bound:
\begin{equation}
\mathbb{E}[\mathcal{P}(\sigma)]=\lim_{N(\mathcal{T}_{c})\rightarrow\infty}\langle\mathcal{P}(
\sigma)\rangle.
\end{equation}
Taking both of the above estimations into account, we then estimate the return probability $\mathbb{E}[\mathcal{P}(\sigma)]$ as
\begin{equation}
\langle\mathcal{P}^{(K)}(\sigma)\rangle=\frac{1}{N(\mathcal{T}_{c})}\sum_{j=1}^{N(\mathcal{T}_{c})}\mathcal{P}_{\mathcal{T}_{c}^{(j)}}^{(K)}(\sigma).
\end{equation}
One clearly recovers the expectation value $\mathbb{E}[\mathcal{P}(\sigma)]$ in the double limit:
\begin{equation}
\mathbb{E}[\mathcal{P}(\sigma)]=\lim_{\substack{K\rightarrow N_{D} \\ N(\mathcal{T})\rightarrow\infty}}\langle\mathcal{P}^{(K)}(\sigma)\rangle.
\end{equation}
We estimate the spectral dimension as
\begin{equation}
\mathcal{D}_{\mathrm{s}}^{(K)}(\sigma)=-2\frac{\mathrm{d}\ln{\langle\mathcal{P}^{(K)}(\sigma)\rangle}}{\mathrm{d}\ln{\sigma}}
\end{equation}
for an appropriate discretization of the derivative with respect to $\sigma$. 


\end{document}